\documentclass[twocolumn,epjc3]{svjour3}

\usepackage{amsmath}
\usepackage{graphicx}
\usepackage{amssymb}
\usepackage{subfigure}
\usepackage{cancel}
\usepackage{xspace}
\usepackage{relsize}
\usepackage{bbold}
\usepackage{lineno}
\usepackage{braket}
\usepackage{slashed}
\usepackage{multirow}
\usepackage{placeins}
\usepackage{rotating}
\usepackage{url}
\usepackage{color}
\usepackage{hyperref}
\usepackage{cite}
\usepackage{lscape}
\usepackage[utf8]{inputenc} 
\usepackage{float}
\usepackage[usenames,dvipsnames,svgnames,table]{xcolor}
\usepackage{colortbl}
\usepackage[all]{hypcap}
\RequirePackage{fix-cm}
\usepackage{soul}

\definecolor{oucrimsonred}{rgb}{0.6, 0.0, 0.0}
\definecolor{persianblue}{rgb}{0.11, 0.22, 0.73}
\definecolor{forestgreen}{rgb}{0.13,0.35,0.13}
\hypersetup{colorlinks, citecolor=oucrimsonred, linkcolor=persianblue, urlcolor=oucrimsonred}

\newcommand{\beq}{\begin{equation}} 
\newcommand{\eeq}{\end{equation}}
\newcommand{\bea}{\begin{eqnarray}}  
\newcommand{\eea}{\end{eqnarray}}
\newcommand{\nnl}{\nonumber \\}

\definecolor{Gray}{gray}{0.9}

\usepackage[T1]{fontenc} 

\smartqed  
\RequirePackage{graphicx}
\journalname{Eur. Phys. J. C}
\begin{document}
\title{\textcolor{black}{Charming Penguins and Lepton Universality Violation in $\boldmath{b \to s \ell^+ \ell^-}$ decays}}
\author{Marco Ciuchini\thanksref{em1,addr1}
\and
Marco Fedele\thanksref{em2,addr2}
\and
Enrico Franco\thanksref{em3,addr3}
\and
Ayan Paul\thanksref{em4,addr5,addr6}
\and
Luca Silvestrini\thanksref{em5,addr3}
\and
Mauro Valli\thanksref{em6,addr7}
}

%
\thankstext{em1}{marco.ciuchini@roma3.infn.it}
\thankstext{em2}{marco.fedele@kit.edu}
\thankstext{em3}{enrico.franco@roma1.infn.it}
\thankstext{em4}{ayan.paul@desy.de}
\thankstext{em5}{luca.silvestrini@roma1.infn.it}
\thankstext{em6}{mauro.valli@stonybrook.edu}
\institute{INFN Sezione di Roma Tre, Via della Vasca Navale 84, I-00146 Rome, Italy \label{addr1}
\and
  Institut f\"ur Theoretische Teilchenphysik, Karlsruhe Institute of Technology, D-76131 Karlsruhe, Germany\label{addr2}
\and
  INFN Sezione di Roma, Piazzale Aldo Moro 2, I-00185 Rome, Italy  \label{addr3}
\and 
Deutsches Elektronen-Synchrotron DESY, Notkestr. 85, 22607 Hamburg, Germany \label{addr5}
\and
Institut f\"ur Physik, Humboldt-Universit\"at zu Berlin, D-12489 Berlin, Germany  \label{addr6}
\and
C.N. Yang Institute for Theoretical Physics, Stony Brook University, Stony Brook, NY 11794,~USA \label{addr7}
}

\date{Received: date / Accepted: date}
\maketitle

\begin{abstract}{
 The LHCb experiment has very recently presented new results on Lepton Universality Violation (LUV) in $B \to K^{(*)} \ell^+ \ell^-$ decays involving $K_S$ in the final state, which strengthen the recent evidence of LUV obtained in $B^+ \to K^{+} \ell^+ \ell^-$ decays and the previous measurements of $B \to K^{*0} \ell^+ \ell^-$. While LUV observables in the Standard Model are theoretically clean, their predictions in New Physics scenarios are sensitive to the details of the hadronic dynamics, and in particular of the charming penguin contribution. In this work, we show how a conservative treatment of hadronic uncertainties is crucial not only to assess the significance of deviations from the Standard Model but also to obtain a \textcolor{black}{conservative} picture of the New Physics responsible for LUV. Adopting a very general parameterization of charming penguins, we find that: \textit{i)} current data hint at a sizable $q^2$ and helicity dependence of charm loop amplitudes;  \textit{ii)} \textcolor{black}{conservative} NP solutions to $B$ anomalies favour a left-handed or an axial lepton coupling rather than a vector one.}
\end{abstract}
\flushbottom


\maketitle

\section{Introduction}
\label{sec:Intro}

Recently, the LHCb experiment at the Large Hadron Collider has announced evidence of Lepton Universality Violation (LUV) in the ratio \cite{LHCb:2021trn}
\begin{eqnarray}
    R_K[1.1,6]& \ \equiv \ & \frac{\mathrm{BR}(B^+ \to K^+ \mu^+ \mu^-)}{\mathrm{BR}(B^+ \to K^+ e^+ e^-)}\vert_{q^2 \in [1.1,6] \mathrm{GeV}^2}    \nonumber \\ 
    & \ \ = \ \ & 0.846^{+0.042\,+0.013}_{-0.039\,-0.012} \,, \label{RKnew}
\end{eqnarray}
crowning with success a huge experimental effort aimed at detecting deviations from the Standard Model (SM) in rare $B_q$ decays. Very recently, another
piece was added to the already very rich set of data on (semi)leptonic and radiative $B_q$ decays: the measurements of \cite{LHCb:2021lvy}
\begin{eqnarray}
    R_{K_S}[1.1,6] & \ \equiv \ & \frac{\mathrm{BR}(B_d \to K_S \mu^+ \mu^-)}{\mathrm{BR}(B_d \to K_S e^+ e^-)}\vert_{q^2 \in [1.1,6] \mathrm{GeV}^2}   \nonumber \\ 
    & \ \ = \ \  & 0.66^{+0.20\,+0.02}_{-0.14\,-0.04} \,, \label{RKShort} \\
    R_{K^{*+}}[0.045,6] & \ \equiv \ & \frac{\mathrm{BR}(B^+ \to K^{*+} \mu^+ \mu^-)}{\mathrm{BR}(B^+ \to K^{*+} e^+ e^-)}\vert_{q^2 \in [0.045,6] \mathrm{GeV}^2} \nonumber \\ 
    & \ \ = \ \  & 0.70^{+0.18\,+0.03}_{-0.13\,-0.04} \,, \label{RKStarp}
\end{eqnarray}
complementing the analogous search for LUV in $B \to K^* \ell^+ \ell^-$ decays, $R_{K^*}$~\cite{Aaij:2017vbb,Abdesselam:2019wac}, the measurement of
BR$(B_s \to \mu^+ \mu^-)$~\cite{CMS:2014xfa,LHCb:2017rmj,ATLAS:2018cur,CMS:2019bbr,LHCb:2021vsc}, the angular analyses and BR measurements of
$B \to K^{(*)} \mu^+ \mu^-$ \cite{Aaij:2020ruw,Aaij:2020nrf,Aaij:2015oid}, $B_s \to \phi \mu^+ \mu^-$~\cite{LHCb:2013tgx,LHCb:2015wdu,LHCb:2021zwz,LHCb:2021xxq} and $B \to K^* e^+ e^-$~\cite{Belle:2016fev,Belle:2019oag}. While hadronic uncertainties
make the detection of possible New Physics (NP) contributions to $B_q \to K^{(*)} \ell^+ \ell^-$ differential rates very difficult, at least with current data, any observation of LUV beyond the percent level due to QED corrections~\cite{Bordone:2016gaq,Isidori:2020acz} would be a clean signal of NP.

\begin{figure}[t!]
  \centering
  \includegraphics[width=\columnwidth]{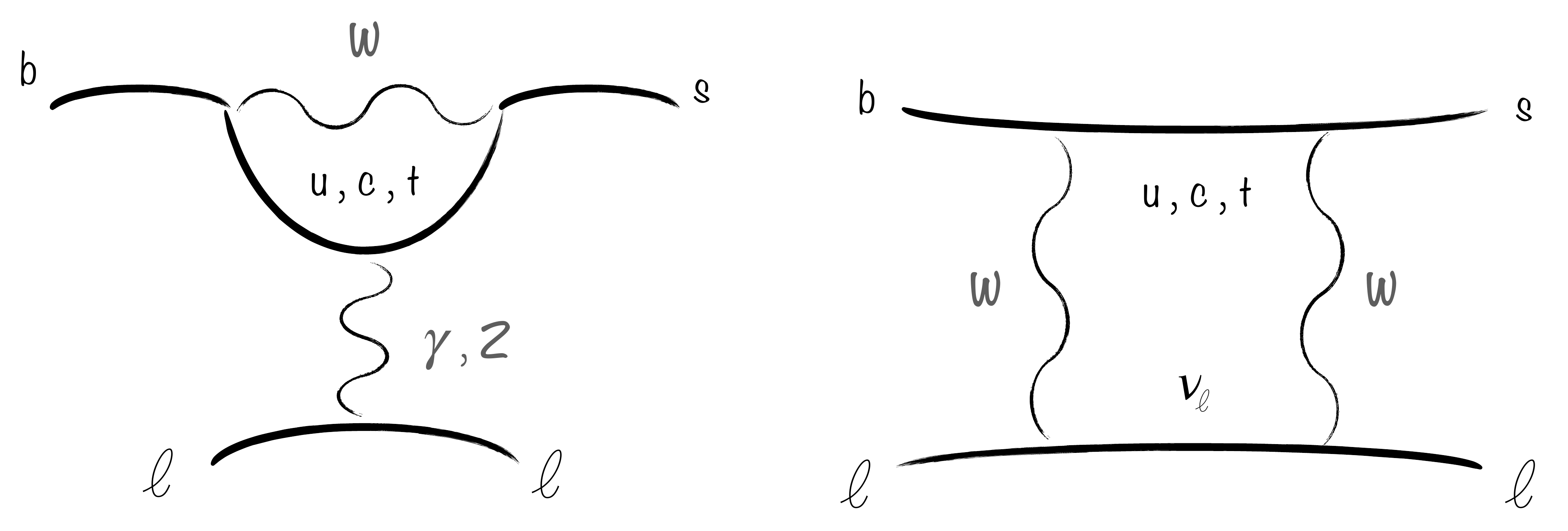}
  \caption{Penguin and box diagrams giving rise to $b \to s \ell^+ \ell^-$ transitions in the SM.}
  \label{fig:diagrams}
\end{figure}

In the SM, $b \to s \ell^+ \ell^-$ transitions can only arise at the loop level, as all other Flavour Changing Neutral Current (FCNC) processes,
and thus they are particularly sensitive to NP. The leading diagrams giving rise to these transitions are illustrated in Fig.~\ref{fig:diagrams}.
Given the hierarchy in the CKM angles, one has $V_{ub}^{} V_{us}^* \ll V_{cb}^{} V_{cs}^* \sim V_{tb}^{} V_{ts}^*$, making the contribution of virtual up-type quarks in the loops negligible.
The GIM mechanism works differently for different diagrams: $Z$-penguins and boxes vanish as $m_q^2/m_W^2$ and are therefore dominated by the top quark, while photonic penguins
have a logarithmic dependence on the quark mass, allowing for a large contribution by the charm quark.

Another fundamental difference between the two classes of diagrams is due
to the chirality of the weak couplings: $Z$-penguins and boxes involve both vector and axial couplings to leptons, while photon penguins couple vectorially to leptons.
This implies that the top-dominated $Z$-penguins and boxes give rise to the local operators $Q_{9V} \sim \bar{b} \gamma^\mu P_L s \bar{\ell} \gamma_\mu \ell$ and
$Q_{10A} \sim \bar{b} \gamma^\mu P_L s  \bar{\ell} \gamma_\mu \gamma_5 \ell$ at the electroweak scale. Photonic penguins instead are more complicated: the top
quark contributes to $Q_{9V}$ at the electroweak scale, but the charm quark remains \textcolor{black}{dynamical} at the scale $m_b$ and therefore contributes to $b \to s \ell^+ \ell^-$ transitions both via
the local operator $Q_{9V}$ and via the (potentially nonlocal and nonperturbative) matrix elements of current-current operators involving the charm quark,
$Q_{1,2}^{\bar{b}c\bar{c}s} \sim \bar{b} \gamma^\mu P_L c \bar{c} \gamma_\mu P_L s$, denoted by charming penguins \cite{Ciuchini:1997hb,Ciuchini:1997rj,Ciuchini:2001gv}. This complication, however, does not affect axial lepton couplings, which remain purely
short-distance.

Computing the matrix element of $Q_{1,2}^{\bar{b}c\bar{c}s}$ is a formidable task.
While the calculation of decay amplitudes for exclusive $b \to s \ell^+ \ell^-$ transitions is well-defined in the infinite $b$ and $c$ mass limit~\cite{hep-ph/9905312,Beneke:2000ry,Beneke:2004dp}, and 
while in the same limit the uncertainty from decay form factors can be eliminated by taking suitable ratios of observables~\cite{Descotes-Genon:2013vna,Descotes-Genon:2014uoa}, in the real world amplitude calculations must 
cope with power corrections~\cite{Jager:2012uw,Jager:2014rwa}, which can be sizable or even dominant in several kinematic regions~\cite{Ciuchini:2015qxb,Ciuchini:2016weo,Arbey:2018ics,Ciuchini:2018anp,Hurth:2020rzx}. For example,
the Operator Product Expansion is known to fail altogether for resonant $B \to K^{(*)} J/\psi \to K^{(*)} \mu^+ \mu^-$ transitions~\cite{Beneke:2009az}, and its accuracy is questionable close to the $c \bar c$ threshold. 
For this reason, estimating corrections to QCD factorization (QCDF) in the low dilepton invariant mass (low-$q^2$) region of $B \to K^{(*)} \ell^+ \ell^-$ and $B_s \to \phi \ell^+ \ell^-$ decay amplitudes is a crucial step towards a
reliable assessment of possible deviations from SM predictions in these decay channels. 
Unfortunately, first-principle calculations of these power corrections are not currently available, and a theoretical breakthrough would be needed to perform such calculations, see, e.g., the discussion
in~\cite{Jager:2014rwa,Bobeth:2017vxj,Melikhov:2019esw}. 

Therefore, as of now, a conservative analysis of semileptonic $B$ decays can only rely on the use of data-driven methods to account for the theoretical uncertainties and to quantify possible deviations from the SM. In this regard, it is important to stress that while the contribution of $Q_{9V}$ to the decay amplitude should depend on helicity and on $q^2$ according to the form factors, long-distance contributions from the charm loop matrix element should show some additional helicity and $q^2$ dependence. It is then very interesting to use a $q^2$ and helicity-dependent parameterization of power corrections~\cite{Jager:2012uw,Jager:2014rwa} when analyzing experimental data: A sizable deviation from
what expected from purely local matrix elements would be a clear confirmation of the presence of power corrections. 

Obviously, the charm loop matrix element cannot generate any Lepton Universality Violation (LUV), so that
ratios of decay Branching Ratios (BRs) for different leptons in the final state can be very reliably predicted in the SM~\cite{Hiller:2003js,Bobeth:2007dw,Bordone:2016gaq,Isidori:2020acz}. However, once lepton non-universal NP is introduced, the hadronic uncertainty
related to the charm loop creeps back in, due to the interference between SM and NP contributions in decay amplitudes, so that the inference of NP parameters from LUV observables is not free from hadronic uncertainties.
Thus, the relevance of a careful and conservative treatment of hadronic uncertainties in assessing the compatibility of $b \to s \ell^+ \ell^-$ data with the SM, and in inferring what kind of NP could lie behind the evidence of LUV, cannot be overstated.

\textcolor{black}{Several analyses~\cite{Hiller:2021pul,Geng:2021nhg,Cornella:2021sby,Hurth:2021nsi,Alguero:2021anc,Bause:2021ply} have recently discussed the implications for NP with a particular focus on the ``clean'' observables: LUV measurements and BR$(B_{s} \to \mu^{+} \mu^{-})$.} 
The aforementioned ``clean'' observables have also been considered together with other interesting tensions with the SM, see for instance~\cite{Crivellin:2020oup,Crivellin:2021njn,Crivellin:2021rbf} for the case of the so-called ``Cabibbo anomaly'', and refs.~\cite{Arnan:2019uhr,Datta:2019bzu,Greljo:2021xmg,Arcadi:2021cwg,Marzocca:2021azj,Perez:2021ddi,Darme:2021qzw,Greljo:2021npi} for possible connections with the long-standing puzzle of the magnetic dipole moment of the muon.

In this work, we focus on $b \to s \ell^+ \ell^-$ transitions in a bottom-up perspective. Building on our previous analyses \cite{Ciuchini:2015qxb,Ciuchini:2016weo,Ciuchini:2017mik,Ciuchini:2018anp,Ciuchini:2019usw,Ciuchini:2020gvn} and on the data presented in~\cite{LHCb:2021trn,LHCb:2021lvy,Aaij:2017vbb,Abdesselam:2019wac,CMS:2014xfa,LHCb:2017rmj,ATLAS:2018cur,CMS:2019bbr,LHCb:2021vsc,Aaij:2020ruw,Aaij:2020nrf,Aaij:2015oid,LHCb:2013tgx,LHCb:2015wdu,LHCb:2021zwz,LHCb:2021xxq,Belle:2016fev,Belle:2019oag},
we aim at answering two fundamental, and deeply related, questions:
\begin{enumerate}
\item Do current data on differential BRs display a non-trivial $q^2$ and helicity dependence of charming penguins, pointing to sizable long-distance effects?  
\item What is the overall significance for NP in light of the new data, and how do hadronic uncertainties affect the interpretation of the present evidence of LUV?
\end{enumerate}

To this end, following the strategy we originally proposed in ref.~\cite{Ciuchini:2015qxb}, in the following we consider a generic parameterization for non-factorizable QCD power corrections, and let data determine the $q^2$ and helicity dependence of charming penguins. 
\textcolor{black}{In particular, in our analysis we determine short-distance and long-distance contributions in a simultaneous fashion exploiting the constraining power of the up-to-date experimental information on $b \to s \ell^{+} \ell^{-}$ together with the current theoretical knowledge of hadronic correlators and matrix elements for this class of rare $B$ meson decays.}

Notice that the approach recently followed in ref.~\cite{Isidori:2021vtc} to allow for an arbitrary lepton-universal correction $\Delta C_9^U$ to $C_9$ is less general, and therefore less conservative, than our approach, unless $\Delta C_9^U$ is promoted from a parameter to a $q^2$- and helicity-dependent function.

The paper is organized as follows: in \autoref{sec:Hadronic} we present our parameterization for charming penguins and discuss the implications of current data on QCD long-distance effects; in \autoref{sec:NP} we present 
a global analysis of NP effects using both the Standard Model Effective Field Theory (SMEFT) and the Weak Effective Hamiltonian; in \autoref{sec:Concl} we wrap up the present study with our conclusions.

\section{Charming penguins from current data}
\label{sec:Hadronic}

For the convenience of the reader, let us briefly summarize our approach to hadronic uncertainties. We write down the helicity-dependent SM decay amplitudes for $B \to K^* \ell^+ \ell^-$ in the following way \cite{Jager:2012uw,Gratrex:2015hna}:
\begin{eqnarray}
H_V^{\lambda} &\ \ \propto\ \ & \left\{C_9^{\rm SM}\widetilde{V}_{L\lambda} + \frac{m_B^2}{q^2} \left[\frac{2m_b}{m_B}C_7^{\rm SM}\widetilde{T}_{L\lambda}  - 16\pi^2h_{\lambda} \right]\right\}\,,\nonumber\\
H_A^{\lambda} &\ \ \propto\ \ & C_{10}^{\rm SM}\widetilde{V}_{L\lambda} \ , \
H_P \propto \frac{m_{\ell} \, m_{b}}{q^2} \,  C_{\rm 10}^{\rm SM} \left( \widetilde{S}_{L} - \frac{m_s}{m_b}\widetilde{S}_{R} \right) \label{Hp}
\label{eq:helamp}
\end{eqnarray}
with $\lambda=0,\pm$ and $C_{7,9,10}^{\rm SM}$ the SM Wilson coefficients of the operators $Q_{9V}$, $Q_{10A}$ and $Q_{7\gamma} \sim  m_b \bar{b}_R\sigma_{\mu\nu}F^{\mu\nu}s_L$ normalized as
in ref.~\cite{Ciuchini:2019usw}.

The factorizable part of the amplitudes corresponds to seven independent form 
factors, $\widetilde{V}_{0,\pm}$, $\widetilde{T}_{0,\pm}$ and $\widetilde{S}$, smooth functions of $q^{2}$~\cite{Straub:2015ica,Gubernari:2018wyi}. Instead, $h_\lambda(q^2)$ represents the non-factorizable part of the amplitude~\cite{Jager:2014rwa,Ciuchini:2015qxb,Chobanova:2017ghn}, dominated by the charming penguin contribution. Nonperturbative methods working in Euclidean spacetime such as lattice QCD cannot directly evaluate $h_\lambda(q^2)$ as at present there is no way to evade the Maiani-Testa no-go theorem~\cite{Maiani:1990ca}, which prevents the computation of rescattering and final state interactions away from the threshold. {In particular, rescattering from an intermediate $D_s^{(*)}-\bar{D}^{(*)}$ state, or from any other on-shell state with flavour content $c \bar{c} \bar{s} d$, into a $K^{(*)}$ and a (virtual) photon is currently not computable. Notice that this intermediate state is always kinematically accessible since $m_{B_d} > m_D^{(*)} + m_{D_s}^{(*)}$, irrespective of the dilepton invariant mass. Such contribution is intrinsically nonperturbative (it could become computable, and negligible, in the $m_b \gg m_c$ limit) and could give a large imaginary part to the charming penguin already at low $q^2$. Another kind of charming penguin corresponds instead to the decay of the $B$ meson into a $K^{(*)}$ and a virtual $c \bar c$ pair, which decays into the lepton pair through the exchange of a photon. This class of charming penguins, giving rise to singularities in $q^2$, can be estimated using light-cone sum rules for $q^2 \ll m_c^2$ or for Euclidean $q^2$~\cite{Khodjamirian:2010vf,Bobeth:2017vxj,Gubernari:2020eft}. To summarize, we can only count on a model-dependent estimate of a subclass of charming penguins in a limited physical region of small $q^2$.} 

Given our ignorance of the charming penguin amplitude, we parameterize the hadronic contribution as follows~\cite{Ciuchini:2018anp}:\textcolor{black}{
\begin{eqnarray} 
\label{eq:newhs}
h_-(q^2) &\ \ =\ \ & -\frac{m_b}{8\pi^2 m_B} \widetilde T_{L -}(q^2) h_-^{(0)} -\frac{\widetilde V_{L -}(q^2)}{16\pi^2 m_B^2} h_-^{(1)} q^2 \nonumber\\
&& + h_-^{(2)} q^4+{\cal O}(q^6)\,, \nonumber\\
h_+(q^2) &=&  -\frac{m_b}{8\pi^2 m_B} \widetilde T_{L +}(q^2) h_-^{(0)} -\frac{\widetilde V_{L +}(q^2)}{16\pi^2 m_B^2}  h_-^{(1)} q^2 \nonumber\\
&& + h_+^{(0)} + h_+^{(1)}q^2 + h_+^{(2)} q^4+{\cal O}(q^6)\,, \nonumber\\
h_0(q^2) &=& -\frac{m_b}{8\pi^2 m_B} \widetilde T_{L 0}(q^2) h_-^{(0)} -\frac{\widetilde V_{L 0}(q^2)}{16\pi^2 m_B^2}  h_-^{(1)} q^2 \nonumber\\
&& + h_0^{(0)}\sqrt{q^2} + h_0^{(1)}(q^2)^\frac{3}{2}  +{\cal O}((q^2)^\frac{5}{2})\,,
\end{eqnarray}
which allows us to write the helicity amplitudes as:}
\begin{eqnarray} 
\label{eq:hv}
H_V^{-} \propto  
 & \ \ \frac{m_B^2}{q^2} \ \ &  \bigg[ \frac{2m_b}{m_B}\left(C_7^{\rm SM} + h_-^{(0)} \right) \widetilde T_{L -}  
  -  16\pi^2 h_-^{(2)}\, q^4 \bigg] \nonumber \\
  & \ \ + \ \ & \left(C_9^{\rm SM} + h_-^{(1)}\right)\widetilde V_{L -}\,, \nonumber \\ H_V^{+} \propto  
 & \ \ \frac{m_B^2}{q^2} & \ \ \bigg[ \frac{2m_b}{m_B}\left(C_7^{\rm SM} + h_-^{(0)} \right) \widetilde T_{L +}  
  -  16\pi^2 \Big(h_{+}^{(0)} 
   \nonumber \\
  & \ \ + \ \ &  h_{+}^{(1)}\, q^2 + h_{+}^{(2)}\, q^4\Big) \bigg] 
  + \left(C_9^{\rm SM} + h_-^{(1)}\right)\widetilde V_{L +}\,, \nnl 
H_V^{0} \propto  
 & \frac{m_B^2}{q^2}&  \bigg[ \frac{2m_b}{m_B}\left(C_7^{\rm SM} + h_-^{(0)} \right) \widetilde T_{L 0}  \nonumber \\
  & \ \ - \ \ &  16\pi^2 \sqrt{q^2} \Big(h_{0}^{(0)}
  + h_{0}^{(1)}\, q^2 \Big) \bigg] \nonumber \\
  & \ \ + \ \ & \left(C_9^{\rm SM} + h_-^{(1)}\right)\widetilde V_{L 0}\,.
  \label{eq:HVs}
\end{eqnarray}

\textcolor{black}{The parameterization in eq.~\eqref{eq:newhs} is a variation of a simple Taylor expansion formally in $q^2/(4m_{c}^2)$, see~\cite{Jager:2014rwa}, whose radius of convergence (for the leading contribution expected from the $c \bar{c}$ loop) is up to the $J/\Psi$ resonance. While this is clearly not the only possible way to parameterize the hadronic contribution, such a choice has been originally introduced in ref.~\cite{Ciuchini:2018anp} with two specific goals in mind. First, from eq.~\eqref{eq:hv} above it is evident that $h_-^{(0)}$ is equivalent to a shift in $C_{7}$, i.e. $\Delta C_7$, while $h_-^{(1)}$ corresponds to a lepton universal correction $\Delta C_9$. Second, the remaining $h$ parameters appearing in \eqref{eq:HVs} are not equivalent to a shift in the Wilson coefficients of $Q_{7\gamma,9V}$ and thus they represent genuine hadronic effects.} {One could take into account the presence of $c \bar{c}$ resonances by adding the corresponding poles to the parameterization. However, when Taylor expanding the full parameterization around $q^2 = 0$, this would just amount to a redefinition of the $h_{\lambda}^{(i)}$ coefficients.}\footnote{Note that our expansion parameter becomes $\mathcal{O}(1)$ in the $q^2$ bin [6,8]~GeV$^{\,2}$. Here we work under the approximation that higher-order terms do not matter in the hadronic ansatz adopted. Any possible relevance of the neglected higher-order terms would only strengthen the conclusion that large hadronic effects are phenomenologically relevant in the analysis.}

As discussed in detail in~\cite{Khodjamirian:2010vf,Ciuchini:2015qxb}, the hadronic contributions introduced above correspond to the following $q^2$- and helicity-dependent shifts in $C_{9}$:
\begin{eqnarray}
\Delta C_{9,1}(q^2) & \ \ = \ \ & - \frac{16 m_{B}^3(m_B+m_{K^{*}})\pi^2}{\sqrt{\lambda(q^2)}V(q^2) q^2} (h_{-}(q^2)-h_{+}(q^2))\nonumber\\
\Delta C_{9,2}(q^2)  & \ \ = \ \ &  - \frac{16 m_{B}^3\pi^2}{(m_B+m_{K^{*}})A_{1}(q^2)q^2} (h_{-}(q^2)+h_{+}(q^2)) \nonumber \\
\Delta C_{9,3}(q^2) & \ \ = \ \ & 
\frac{64 \pi^2 m_B^3 m_{K^{*}} \sqrt{q^2} (m_B + m_{K^*})}{\lambda(q^2) A_{2}(q^2) q^2} \, h_{0}(q^2) \nonumber \\ 
& \ - \ &\frac{16 m_{B}^3(m_B+m_{K^{*}})(m_B^2-q^2-m_{K^{*}}^2)\pi^2}{\lambda(q^2) A_{2}(q^2) q^2} \nonumber \\  
 & \ \times \ & (h_{-}(q^2)+h_{+}(q^2)) \ .
\label{eq:gtildes}
\end{eqnarray}

\begin{figure*}[t!]
  \centering
  \includegraphics[width=0.8\textwidth]{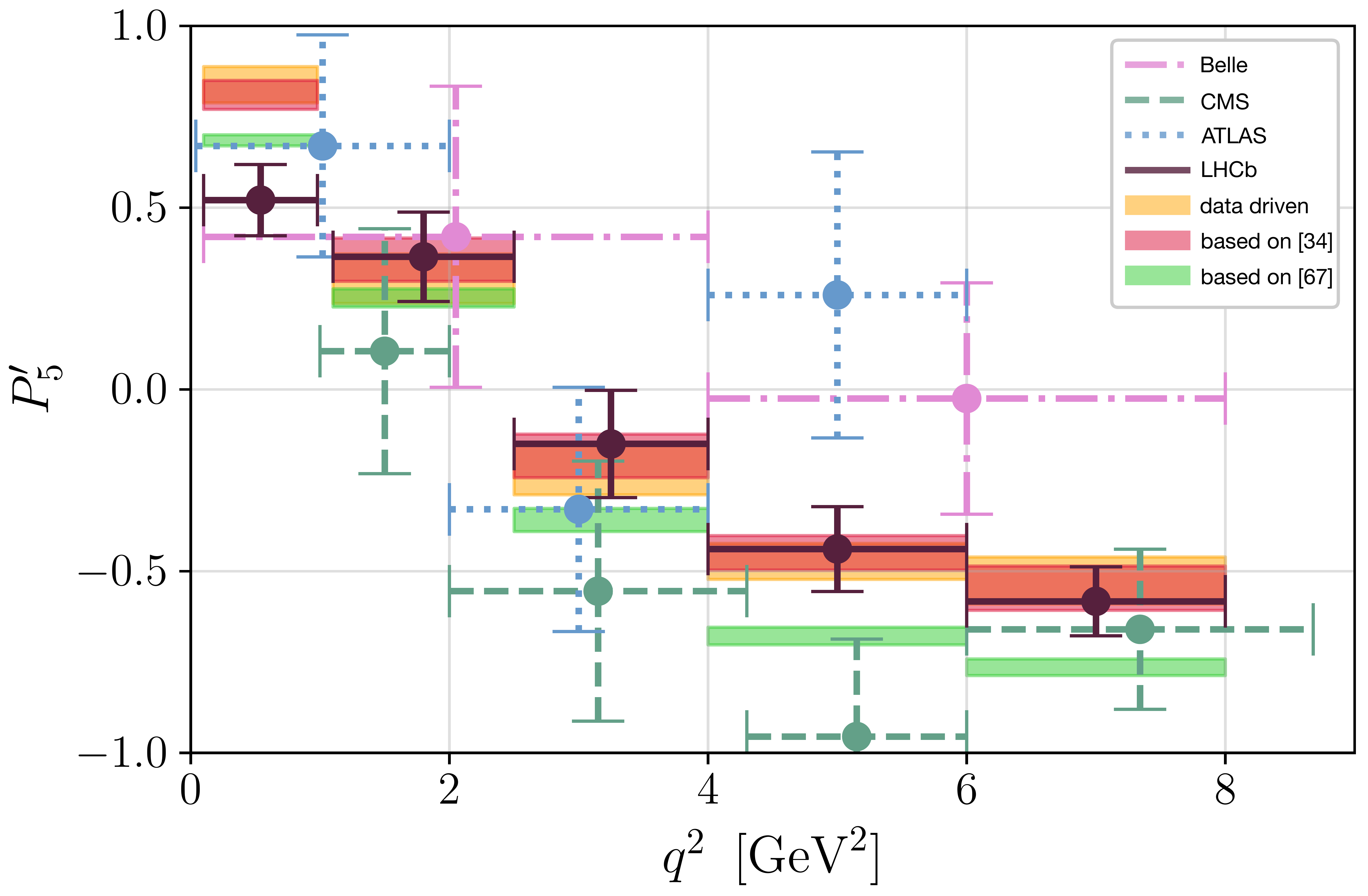}
  \caption{The result for the optimized observable $P_5^\prime$ (see ref.~\cite{DescotesGenon:2012zf}) from a global fit within the SM in three different approaches for the charming penguin contribution, compared to experimental data. Only the approach fully relying on LCSR estimates (refs.~\cite{Khodjamirian:2010vf,Gubernari:2020eft}) and their 
  \textcolor{black}{extension} by means of dispersion relations in the entire dilepton invariant-mass range may lead to the so-called ``$P_{5}^{\prime}$ anomaly''. 
}
  \label{fig:P5p}
\end{figure*}

An analogous parameterization can be introduced for charming penguins in $B \to K \ell^+ \ell^-$, as well as for $B_s \to \phi \ell^+ \ell^-$. When considering hadronic contributions in $B\to K^*$ decays versus those in $B_s\to \phi$ decays the consideration of flavour SU(3) breaking comes into play. Given that the degree of SU(3) breaking originating from $m_s \gg m_d,m_u$ is not quantifiable ab initio, we performed some tests by adding ad hoc SU(3) breaking parameters that multiplicatively modifies the hadronic terms in $B_s\to \phi$ decays vis-\`a-vis those in $B\to K^*$ decays, i.e.,
\begin{equation}
    h_\lambda^{B_s\to \phi} = (1+\delta_R + i\delta_I)h_\lambda^{B\to K^*}.
\end{equation}
The two parameters, $\delta_R$ and $\delta_I$, are taken for simplicity to be independent of helicity and simply modify the real and imaginary parts of the hadronic terms independently. We set a Gaussian prior on $\delta_R$ and $\delta_I$ with $\mu=0$ and $\sigma=0.3$ representative of 30\% SU(3) breaking centered at no SU(3) breaking. The posterior distribution of $\delta_R$ and $\delta_I$ are not significantly different from the prior distributions leading us to conclude that the experimental results are not precise enough to draw conclusion about $SU(3)$ breaking within this hypothesis. For the rest of the discussion we assume exact flavour $SU(3)$ symmetry for power corrections, which is justified given the current experimental uncertainties. In the future, realistic departures from the $SU(3)$ limit could be probed by data and the present investigation of QCD effects could be further generalized along these lines in a straightforward manner.

\begin{figure}[ht!]
  \centering
  \includegraphics[width=0.45\textwidth]{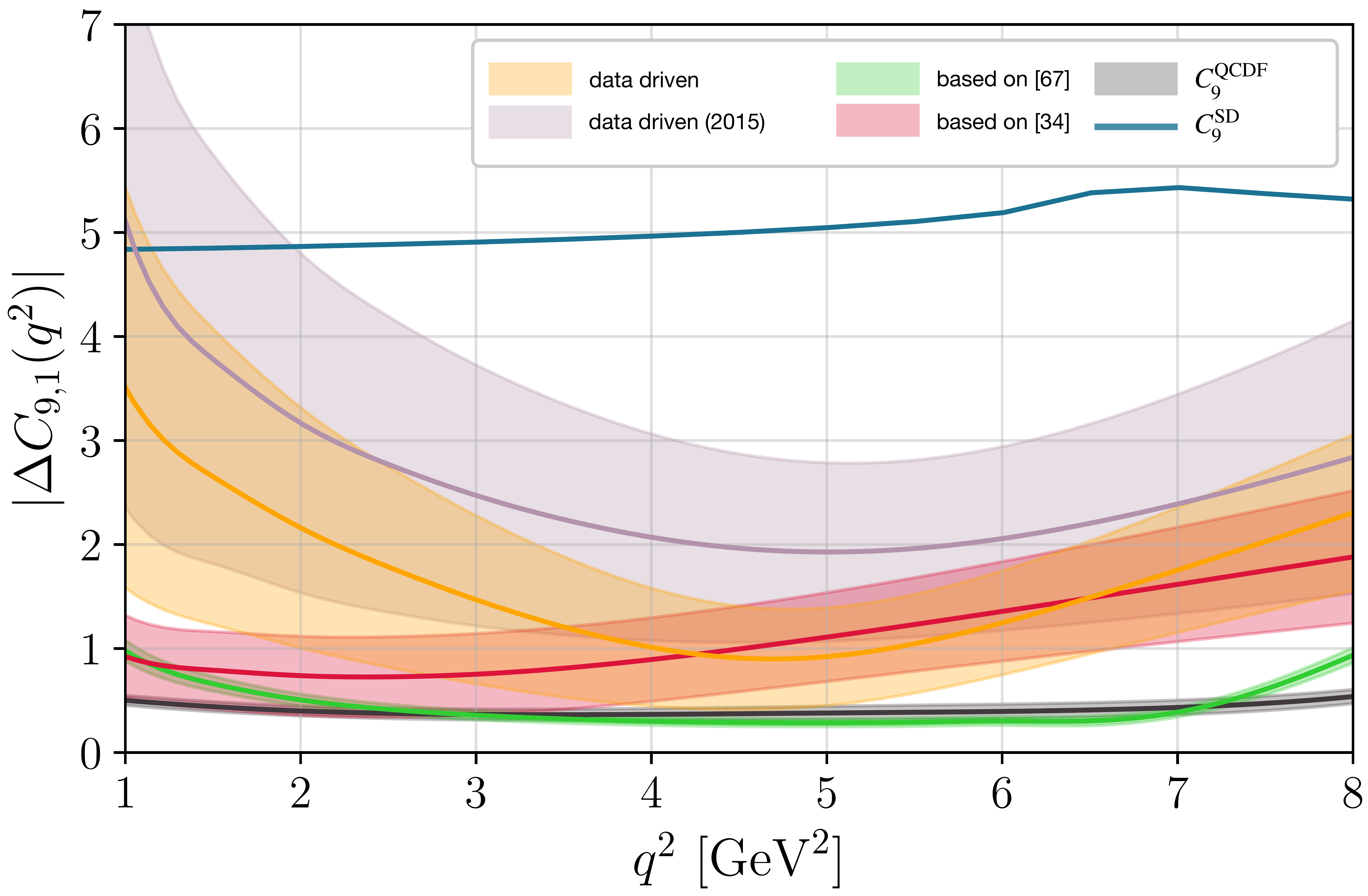}
  \includegraphics[width=0.45\textwidth]{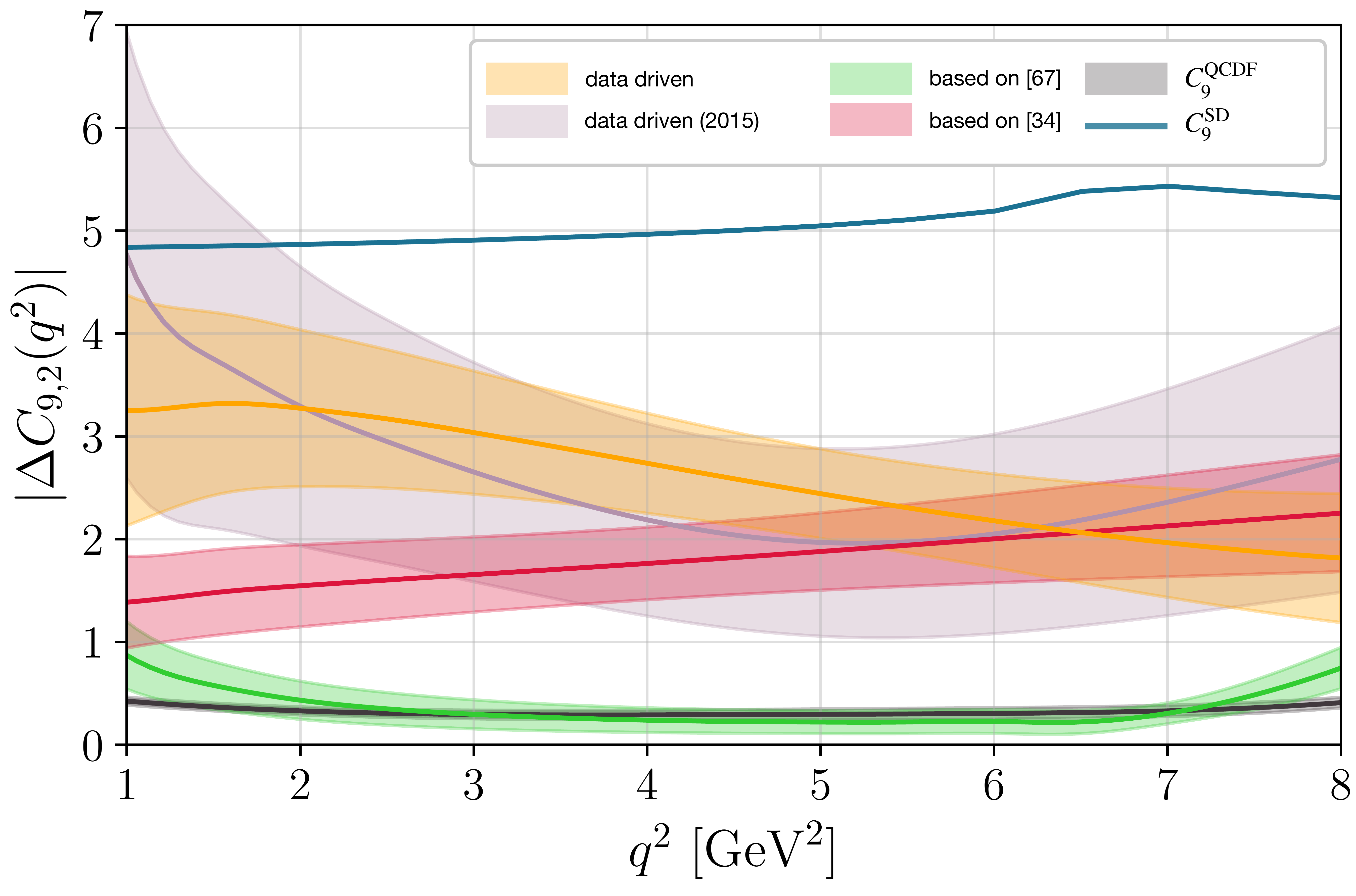}
  \includegraphics[width=0.45\textwidth]{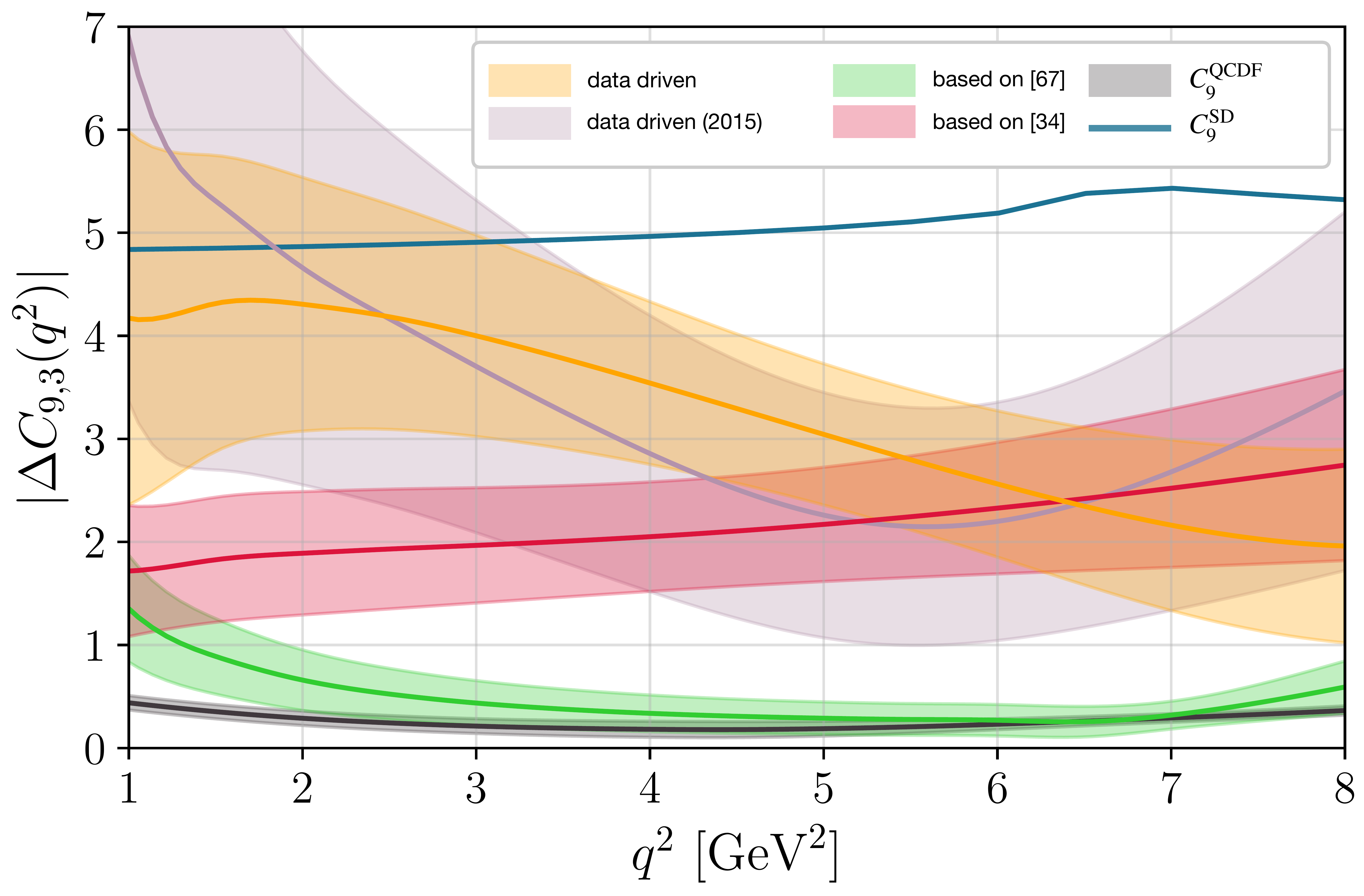}
  \caption{95\% probability contours of the posteriors for the functions $\Delta C_{9,i}(q^2)$ defined in eq.~(\ref{eq:gtildes}) in the three approaches for charming penguins. For comparison, the result obtained in the data-driven approach with 2015 data is also reported, along with the short-distance contribution and the factorizable QCD corrections. }
  \label{fig:deltaC9}
\end{figure}

Using the \texttt{HEPfit} code~\cite{deBlas:2019okz,HEPfit} and the form factors and input parameters used in Refs.~\cite{Ciuchini:2015qxb,Ciuchini:2016weo,Ciuchini:2017mik,Ciuchini:2018anp,Ciuchini:2019usw,Ciuchini:2020gvn}, we perform a Bayesian fit to the data in refs.~\cite{CMS:2014xfa,LHCb:2017rmj,ATLAS:2018cur,CMS:2019bbr,LHCb:2021vsc,Aaij:2020ruw,Aaij:2020nrf,Aaij:2015oid,LHCb:2013tgx,LHCb:2015wdu,LHCb:2021zwz,LHCb:2021xxq,Belle:2016fev,Belle:2019oag} within the SM, adopting three distinct approaches to account for power corrections to QCDF:
\begin{enumerate}
    \item A {model-dependent} approach in which LCSR results are 
    \textcolor{black}{extended} over the full range of $q^2$ {taking analiticity constraints into account (in particular, the singularities of the amplitudes in $q^2$)}~\cite{Khodjamirian:2010vf,Blake:2017fyh,Bobeth:2017vxj,Chrzaszcz:2018yza,Gubernari:2020eft}, where the results either stem directly from a LCSR computation performed at $q^2 \leq 1$ GeV$^2$, or from an extension of these results in the whole phenomenological region by means of dispersion relations. Within this approach, {the parameterization in eq.~\eqref{eq:HVs} is replaced by the ones in eqs.~(7.14) and (7.7) of ref.~\cite{Khodjamirian:2010vf} for $K^*$ and $K$ respectively. In this case, the size of charming penguins in $B \to K^{*} \ell^{+}\ell^{-} $ is comparable to the ones provided by QCDF $\mathcal{O}(\alpha_s)$ corrections $C^{\rm QCDF}_{9}$}, while the size of charming penguins in $B \to K \ell^{+}\ell^{-} $ is too small to be phenomenologically relevant.
    \item A model-dependent approach in which \textcolor{black}{the LCSR results of ref.~\cite{Khodjamirian:2010vf} are used to constrain the $h_{\lambda}$-parameters in eq.~\eqref{eq:newhs}} only for $q^2 \leq 1$ GeV$^2$ (with no specific structure assumed in relation to the singularities in $q^2$).
    \textcolor{black}{Using the results of ref.~\cite{Gubernari:2020eft} instead of ref.~\cite{Khodjamirian:2010vf} would imply an even smaller hadronic contribution at $q^2 \leq 1~$GeV$^2$.}
    Within this approach, the size of charming penguins in $ B \to K^{*} \ell^{+}\ell^{-} $ can depart from the LCSR estimate once away from the light-cone region, while charm-loop effects in $B \to K \ell^{+}\ell^{-} $ are still regarded as negligible in light of the estimates in refs.~\cite{Khodjamirian:2010vf,Gubernari:2020eft}.
    \item A data-driven approach in which all $h$ parameters are determined from experimental data.
    Here, the size of non-perturbative charming penguins can be comparable to the short-distance contributions $C^{\rm SD}_{9}$ for all decay channels. Such an approach can be motivated by further contributions from intermediate-state rescattering like $D_{s}^{(*)}-\bar{D}^{(*)}$ pairs, for which no theoretical estimate is currently available.
\end{enumerate}

In  Fig.~\ref{fig:P5p}, as a representative example, we show the outcome of the fit in the SM for the optimized observable $P^{\prime}_{5}$~\cite{DescotesGenon:2012zf} in the three different approaches to power corrections. As shown in the plot, no tension from data emerges from this observable in the data-driven approach or in the one from ref.~\cite{Ciuchini:2018anp}.
In general, as already noted in~\cite{Ciuchini:2015qxb}, an excellent fit to the data (except, of course, for LUV ratios and with the notable other exception of the time-integrated BR$(B_{s} \to \mu^{+} \mu^{-})$) can be obtained within these two approaches. On the other hand, LCSR results 
\textcolor{black}{extended} to larger values of $q^2$ yield a poor fit of several BRs as well as tension in some of the angular observables, giving rise to the so-called $P_5^\prime$ anomaly. 

\textcolor{black}{We note that for $q^2 \to 0$, $P'_5$ is  only sensitive to the combinations $ \Delta C_{9,1 \pm 2}(q^2)$. Also, $ \Delta C_{9,1 + 2}(q^2)$ rapidly vanishes towards $q^2=0$ in our model-dependent approach based on ref.~\cite{Khodjamirian:2010vf}. This is the result of helicity suppression~\cite{Jager:2012uw}, expected to hold on the light cone, i.e. only at $q^2=0$~\cite{Khodjamirian:2010vf}. Therefore, in the model-dependent approach based on ref.~\cite{Ciuchini:2018anp} we enforce such cancellation only at $q^2=0$. Moreover, in the data-driven approach we do not require it to happen. Consequently, in the first bin of $P'_5$ in Fig.~2 the red and orange bands lie above the green one.}

The result shown for $P^{\prime}_{5}$ highlights the major role played by long-distance effects.
In Fig.~\ref{fig:deltaC9}, we further investigate this aspect showing the $q^2$ and helicity dependence of the charming penguin contributions. In the plot we show the 95\% probability regions of the posteriors for the functions $\Delta C_{9,i}(q^2)$ obtained in the global fit in the SM under the three different approaches. In the same figure, as a guideline, we also show the size of the SM short-distance contribution to $Q_{9V}$, labeled by $C^{\rm SD}_{9}$, as well as the size of the factorizable QCD corrections.

The posteriors of $\Delta C_{9,i}(q^2)$ in Fig.~\ref{fig:deltaC9} display non-negligible hadronic contributions -- comparable in size to $C^{\rm SD}_{9}$ rather than $C^{\rm QCDF}_{9}$ --  in the whole region of low dilepton mass probed by current data. This is not surprising since power corrections are naively expected to be larger than perturbative QCD corrections of $\mathcal{O}(\alpha_s/(4 \pi))$ \cite{Ciuchini:1997hb,Ciuchini:1997rj,Ciuchini:2001gv}. A departure from LCSR expectations even at very low $q^2$ is hinted at in the data-driven approach, which matches the outcome of the approach based on ref.~\cite{Ciuchini:2018anp} only for $q^2 \gtrsim 4 m_{c}^2$.

As can also be seen in Fig.~\ref{fig:deltaC9}, by comparing the Data Driven determination of $\Delta C_{9,i}(q^2)$ from current data with the one obtained in our 2015 analysis \cite{Ciuchini:2015qxb}\footnote{The outcome of the 2015 analysis has been rederived adopting the $h$'s parameterization of this work.} it is evident how improved data on differential BR's allow for a much better knowledge of the charm contribution. In this respect, it will be interesting to see whether more precise data will bring stronger evidence of such hadronic effects. 
At present, hints of large hadronic contributions from data are still statistically mild, as can be read from Table \ref{tab:hlambda}. There, we report the highest probability density intervals (HPDI) for the posteriors of the $h$ parameters adopted in the two more conservative approaches. Some $h$'s corresponding to genuine hadronic contributions deviate from 0, but still only at the $2\sigma$ level. 

\textcolor{black}{To summarize, }the data driven scenario stands out as \textcolor{black}{our} most conservative choice for 
\textcolor{black}{a statistically} robust inference on NP contributions \textcolor{black}{in current $b \to s \ell^{+} \ell^{-}$ data, accounting also for possible large contributions from QCD rescattering}.~\footnote{While large hadronic effects will lead to a conservative estimate of NP in $C_{9}$, in our global analysis -- where $R_{K^{(*)}}$ ratios signal NP irrespective of the hadronic treatment -- such conservative approach will also impact the inference of a non-vanishing NP contribution in $C_{10,\mu,e}$.} In the following we take it as a reference, but for completeness we present results on NP also in the other two approaches.

\begin{table*}[t!]
\centering
\renewcommand{\arraystretch}{1.1}
\begin{tabular}{|c|c|cc|}
\hline
\textbf{Hadronic parameter} & \textbf{Approach} & \textbf{68\% HPDI} & \textbf{95\% HPDI} \\
\hline
\multirow{2}{*}{$ \Re \, h_0^{(0)}  \times 10^4 $ }
& data driven &
$ \color{red} [ 1.69 , 5.83 ]$ &
$[ -0.26 , 8.33 ]$ \\
& based on~\cite{Ciuchini:2018anp} &
$ \color{red} [ 1.91 , 5.25 ]$ &
$ \color{red} [ 0.26 , 7.12 ]$ \\
\hline
\multirow{2}{*}{$ \Im \, h_0^{(0)}  \times 10^4 $ }
& data driven &
$[ -4.56 , 2.76 ]$ &
$[ -8.44 , 6.52 ]$ \\
& based on~\cite{Ciuchini:2018anp} &
$[ -4.30 , 0.21 ]$ &
$[ -6.21 , 2.84 ]$ \\
\hline
\multirow{2}{*}{$ \Re \, h_+^{(0)} \times 10^4 $ }
& data driven &
$ \color{red} [ -1.25 , -0.34 ]$ &
$[ -1.73 , 0.12 ]$ \\
& based on~\cite{Ciuchini:2018anp} &
$[ -0.18 , 0.07 ]$ &
$[ -0.34 , 0.19 ]$ \\
\hline
\multirow{2}{*}{$ \Im \, h_+^{(0)} \times 10^4 $ }
& data driven &
$[ -0.40 , 0.65 ]$ &
$[ -0.92 , 1.22 ]$ \\
& based on~\cite{Ciuchini:2018anp} &
$[ -0.12 , 0.11 ]$ &
$[ -0.26 , 0.25 ]$ \\
\hline
\multirow{2}{*}{$ \Re \, h_-^{(0)} \equiv \Re \, \Delta C_{7} \times 10^2 $ }
& data driven &
$[ -0.30 , 2.98 ]$ &
$[ -2.08 , 4.85 ]$ \\
& based on~\cite{Ciuchini:2018anp} &
$[ -1.06 , 1.41 ]$ &
$[ -2.12 , 2.37 ]$ \\
\hline
\multirow{2}{*}{$ \Im \, h_-^{(0)} \equiv \Im \, \Delta C_{7} \times 10^2 $ }
& data driven &
$[ -8.89 , 1.39 ]$ &
$[ -14.15 , 6.57 ]$ \\
& based on~\cite{Ciuchini:2018anp} &
$ {\color{red} [ -3.27 , -0.66 ]} \cup { \color{red} [ 1.46 , 1.60 ]}$ &
$[ -3.64 , 2.75 ]$ \\
\hline
\multirow{2}{*}{$ \Re \, h_0^{(1)} \times 10^5 $ }
& data driven &
$[ -3.61 , 1.88 ]$ &
$[ -6.52 , 5.09 ]$ \\
& based on~\cite{Ciuchini:2018anp} &
$[ -1.87 , 3.03 ]$ &
$[ -4.26 , 5.95 ]$ \\
\hline
\multirow{2}{*}{$ \Im \, h_0^{(1)} \times 10^5 $ }
& data driven &
$[ -7.59 , 2.26 ]$ &
$[ -11.41 , 7.67 ]$ \\
& based on~\cite{Ciuchini:2018anp} &
$[ -5.04 , 3.82 ]$ &
$[ -8.81 , 8.38 ]$ \\
\hline
\multirow{2}{*}{$ \Re \, h_+^{(1)} \times 10^4 $ }
& data driven &
$ \color{red} [ 1.33 , 2.78 ]$ &
$ \color{red} [ 0.52 , 3.51 ]$ \\
& based on~\cite{Ciuchini:2018anp} &
$ \color{red} [ 0.10 , 0.90 ]$ &
$[ -0.33 , 1.29 ]$\\
\hline
\multirow{2}{*}{$ \Im \, h_+^{(1)} \times 10^4 $ }
& data driven &
$ \color{red} [ 0.77 , 2.64 ]$ &
$[ -0.32 , 3.48 ]$ \\
& based on~\cite{Ciuchini:2018anp} &
$ \color{red} [ 0.03 , 0.89 ]$ &
$[ -0.45 , 1.29 ]$ \\
\hline
\multirow{2}{*}{$ \Re \, h_-^{(1)} \equiv \Re \,  \Delta C_{9} $ }
& data driven &
$[ -0.02 , 1.23 ]$ &
$[ -0.68 , 1.87 ]$ \\
& based on~\cite{Ciuchini:2018anp} &
$[ -0.32 , 0.73 ]$ &
$[ -0.88 , 1.23 ]$ \\
\hline
\multirow{2}{*}{$ \Im \, h_-^{(1)} \equiv \Im \,  \Delta C_{9} $ }
& data driven &
$[ -0.90 , 2.68 ]$ &
$[ -2.69 , 4.58 ]$ \\
& based on~\cite{Ciuchini:2018anp} &
$[ -0.05 , 1.91 ]$ &
$[ -1.40 , 2.47 ]$ \\
\hline
\multirow{2}{*}{$ \Re \, h_+^{(2)} \times 10^5 $ }
& data driven &
$ \color{red} [ -3.61 , -1.09 ]$ &
$[ -4.93 , 0.23 ]$ \\
& based on~\cite{Ciuchini:2018anp} &
$[ -0.99 , 0.71 ]$ &
$[ -1.83 , 1.63 ]$ \\
\hline
\multirow{2}{*}{$ \Im \, h_+^{(2)} \times 10^5 $ }
& data driven &
$ \color{red} [ -4.25 , -1.14 ]$ &
$[ -5.74 , 0.59 ]$ \\
& based on~\cite{Ciuchini:2018anp} &
$[ -1.71 , 0.22 ]$ &
$[ -2.63 , 1.23 ]$ \\
\hline
\multirow{2}{*}{$ \Re \, h_-^{(2)} \times 10^5 $ }
& data driven &
$ \color{red} [ 0.15 , 1.77 ]$ &
$[ -0.70 , 2.65 ]$ \\
& based on~\cite{Ciuchini:2018anp} &
$ \color{red} [ 0.98 , 2.33 ]$ &
$ \color{red} [ 0.29 , 3.05 ]$ \\
\hline
\multirow{2}{*}{$ \Im \, h_-^{(2)} \times 10^5 $ }
& data driven &
$[ -1.96 , 2.10 ]$ &
$[ -4.11 , 4.15 ]$ \\
& based on~\cite{Ciuchini:2018anp} &
$[ -1.40 , 1.30 ]$ &
$[ -2.61 , 2.79 ]$ \\
\hline
\end{tabular}
\caption{68\%  and 95\% HPDI of the posterior distribution of the hadronic parameters $h_\lambda^{(i)}$. The \textcolor{red}{red} color highlights ranges not including 0. Genuine hadronic effects encoded in $h^{(0)}_{0}$ and $h^{(2)}_{-}$ are found to be non-vanishing at the 2$\sigma$ level in the model-dependent approach based on ref.~\cite{Ciuchini:2018anp}, while $h^{(1)}_{+}$ deviates from zero at more than $2\sigma$ in the data driven fit.}
\label{tab:hlambda}
\end{table*}

\section{New Physics in \texorpdfstring{$\boldmath{B}$}{B} decays}
\label{sec:NP}

While experimental data on BR's and angular distributions can be reproduced within the SM in both the data-driven approach and in the model-dependent one based on ref.~\cite{Ciuchini:2018anp}, reproducing the central values of the LUV ratios for $B \to K^{(*)} \ell^+ \ell^-$, as well as the current measurement of BR$(B_{s} \to \mu^{+} \mu^{-}$), undoubtedly requires physics beyond the SM.

Given the bounds from direct searches of NP at the LHC, it is reasonable to assume in this context that NP contributions would arise at energies much larger than the weak scale. Then, a suitable framework to describe such contributions is given by the SMEFT, in particular by adding to the SM the following additional dimension-six operators:\footnote{Notice that these operators may be further generated at one loop via SM RGE effects, see, e.g., refs.~\cite{Celis:2017doq,Alasfar:2020mne}. In addition, here we do not consider the possibility that, integrating out NP, one would generate sizable $Q_{1,2}^{\bar{b}c\bar{c}s}$ as studied e.g.  in~\cite{Jager:2017gal,Jager:2019bgk}.}
\bea \label{eq:SMEFT_op_tree}
O^{LQ^{(1)}}_{2223} & \ \ = \ \ & (\bar{L}_2\gamma_\mu L_2)(\bar{Q}_2\gamma^\mu Q_3)\,, \nonumber \\
O^{LQ^{(3)}}_{2223} & \ \ = \ \ & (\bar{L}_2\gamma_\mu \tau^{A} L_2)(\bar{Q}_2\gamma^\mu\tau^{A} Q_3)\,,\nonumber \\
O^{Qe}_{2322} & \ \ = \ \ & (\bar{Q}_{2}\gamma_\mu Q_{3})(\bar{e}_{2} \gamma^\mu e_{2})\,,\nonumber \\
O^{Ld}_{2223} & \ \ = \ \ & (\bar{L}_2\gamma_\mu L_2)(\bar{d}_2\gamma^\mu d_3)\,,\nonumber \\
O^{ed}_{2223} & \ \ = \ \ & (\bar{e}_2\gamma_\mu e_2)(\bar{d}_2\gamma^\mu d_3)\,,
\eea
where $\tau^{A=1,2,3}$ are Pauli matrices (a sum over $A$ in the equations above is understood), weak doublets are in upper case and $SU(2)_{L}$ singlets are in lower case, and flavour indices are defined in the basis of diagonal down-type quark Yukawa couplings. Since in our analysis operators $O^{LQ^{(1,3)}}_{2223}$ always enter as a sum, we collectively denote their Wilson coefficient as $C^{LQ}_{2223}$. For concreteness, we normalize SMEFT Wilson coefficients to a NP scale $\Lambda_\mathrm{NP} = 30$ TeV and we only consider NP contributions to muons.\footnote{The focus on LUV effects in muons is mainly motivated by the $\sim2.3 \sigma$ tension of the SM with the current experimental average for the time-integrated BR$(B_{s} \to \mu^{+} \mu^{-})$.} Matching the SMEFT operators onto the weak effective Hamiltonian one obtains the following contributions to operators $Q_{9V}$ and $Q_{10A}$ and to the chirality-flipped $Q_{9V}^\prime$ and $Q_{10A}^\prime$ \cite{Aebischer:2015fzz}: 
\bea \label{eq:SMEFT_matching}
C_{9}^{\rm NP} & \ \ = \ \ & \frac{\pi v^2}{\alpha_{e}\lambda_{t}\Lambda_{\rm NP}^2} \, \left(C^{LQ^{(1)}}_{22 23} + C^{LQ^{(3)}}_{22 23} + C^{Qe}_{23 22} \right)\,,\nonumber \\
C_{10}^{\rm NP} & \ \ = \ \ & \frac{\pi v^2}{\alpha_{e}\lambda_{t}\Lambda_{\rm NP}^2} \, \left(C^{Qe}_{23 22} - C^{LQ^{(1)}}_{22 23} - C^{LQ^{(3)}}_{22 23}  \right)\,,\nonumber \\
C^{\prime, \rm{NP}}_{9} & \ \ = \ \ & \frac{\pi v^2}{\alpha_{e}\lambda_{t}\Lambda_{\rm NP}^2} \, \left(C^{ed}_{22 2 3} + C^{Ld}_{22 2 3} \right)\,,\nonumber \\
C^{\prime, \rm{NP}}_{10} & \ \ = \ \ & \frac{\pi v^2}{\alpha_{e}\lambda_{t}\Lambda_{\rm NP}^2} \, \left(C^{ed}_{22 2 3} - C^{Ld}_{22 23} \right)\,,
\eea
with $\alpha_{e}$ the fine-structure constant, $v$ the vacuum expectation value of the SM Higgs field, $\lambda_t= V_{ts}^{}V_{tb}^*$, and alignment in the down-quark sector assumed, i.e. $Q_{i} = (V_{ji}^{*}\,u_{jL}, d_{jL} )^{T}$~\cite{Ciuchini:2019usw}.

\begin{table*}[t!]
{\footnotesize
\centering
\renewcommand{\arraystretch}{2.0}
\begin{tabular}{|l|c|cc|}
\hline
\textbf{NP scenario} & \textbf{Approach} & \textbf{68\% HPDI} & \boldmath$\Delta IC$ \\
\hline
\multirow{3}{*}{A:  $ \ C^{\rm NP}_{9} $ }
& data driven &
$[ -3.04 , -1.10 ] \cup [ 1.48 , 1.99 ]
$ & 21 $\cup$ 13 
\\
& based on~\cite{Ciuchini:2018anp} &
$[ -1.44 , -1.01 ]$ & 43 \\
& based on~\cite{Khodjamirian:2010vf} &
$[ -1.37 , -1.12 ]$ & 94 \\
\hline
\multirow{3}{*}{B:  $ \ C^{LQ}_{2223} $ }
& data driven &
$[ 0.65 , 1.05 ]$ & 38 \\
& based on~\cite{Ciuchini:2018anp} &
$[ 0.67 , 0.88 ]$ & 60 \\
& based on~\cite{Khodjamirian:2010vf} &
$[ 0.77 , 0.96 ]$ & 75 \\
\hline
\multirow{3}{*}{C:  $ \ C_{10}^\mathrm{NP} $ }
& data driven &
$[ 0.53 , 0.79 ]$ & 39 \\
& based on~\cite{Ciuchini:2018anp} &
$[ 0.66 , 0.90 ]$ & 54 \\
& based on~\cite{Khodjamirian:2010vf} &
$[ 0.56 , 0.79 ]$ & 20 \\
\hline
\multirow{3}{*}{D: $ \{C^{LQ}_{2223},C^{Qe}_{2322}\} $}
& data driven &
$\{[ 0.20 , 1.03 ], [ -0.82 , 0.15 ]\}$ & 37 \\
& based on~\cite{Ciuchini:2018anp} &
$\{[ 0.61 , 0.86 ], [ -0.37 , 0.11 ]\}$ & 57 \\
& based on~\cite{Khodjamirian:2010vf} &
$\{[ 0.90 , 1.10 ], [ 0.53 , 0.79 ]\}$ & 96 \\
\hline
\multirow{3}{*}{D: $ \{C^{\rm NP}_{9},C^{\rm NP}_{10}\} $}
& data driven &
$\{[ -0.81 , 0.46 ], [ 0.51 , 0.83 ]\}$ & 37 \\
& based on~\cite{Ciuchini:2018anp} &
$\{[ -0.67 , -0.20 ], [ 0.47 , 0.76 ]\}$ & 57 \\
& based on~\cite{Khodjamirian:2010vf} &
$\{[ -1.33 , -1.06 ], [ 0.15 , 0.34 ]\}$ & 96 \\
\hline
E: $\{C^{LQ}_{2223},C^{Qe}_{2322},$
 & data driven & $\{[ -0.06 , 1.18 ], [ -0.99 , 0.35 ], [ -1.30 , 0.34 ], [ -1.25 , 0.56 ]\}$ & 30 \\
 $ \ \ \ \ \ \ C^{Ld}_{2223},C^{ed}_{2223}\}$ & based on~\cite{Ciuchini:2018anp} &
$\{[ 0.83 , 1.32 ], [ -0.05 , 0.76 ], [ -0.59 , -0.10 ], [ -0.58 , 0.27 ]\}$ & 54 \\
& based on~\cite{Khodjamirian:2010vf} &
$\{[ 1.03 , 1.23 ], [ 0.69 , 0.97 ], [ -0.49 , -0.17 ], [ -0.25 , 0.43 ]\}$ & 105 \\
\hline
E: $\{C^{\rm NP}_{9},C^{\rm NP}_{10},$
 & data driven & $\{[ -1.05 , 0.75 ], [ 0.38 , 0.81 ], [ -0.57 , 1.82 ], [ -0.31 , 0.12 ]\}$ & 30 \\
 $ \ \ \ \ \ \ \ C^{\prime, \rm NP}_{9},C^{\prime, \rm NP}_{10}\}$ & based on~\cite{Ciuchini:2018anp} &
$\{[ -1.45 , -0.59 ], [ 0.29 , 0.70 ], [ -0.06 , 0.82 ], [ -0.37 , 0.08 ]\}$ & 54 \\
& based on~\cite{Khodjamirian:2010vf} &
$\{[ -1.55 , -1.27 ], [ 0.11 , 0.31 ], [ -0.17 , 0.52 ], [ -0.47 , -0.14 ]\}$ & 105 \\
\hline
\end{tabular}
\caption{$68\%$ HPDI of the posterior distribution of the SMEFT Wilson coefficients from a fit to the full set of $b \to s \ell^{+} \ell^{-}$ data in the NP scenarios A, B, C, D and E, along with $\Delta IC \equiv IC_{\textrm{SM}} - IC_{\textrm{NP}}$. 
\label{tab:WC_SMEFT}}}
\end{table*}

\begin{figure}[t!]
\centering
\includegraphics[width=0.45\textwidth]{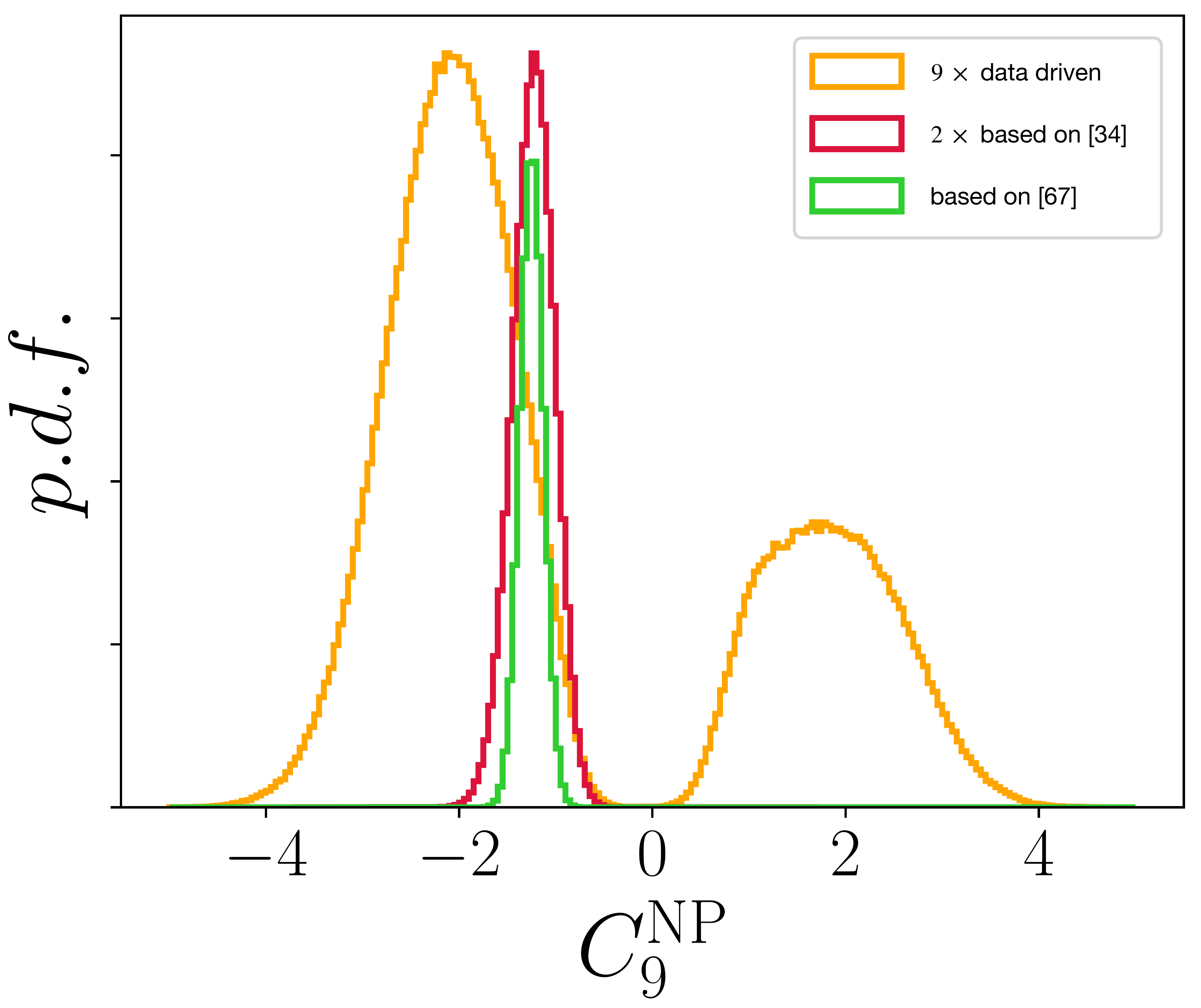}
\includegraphics[width=0.45\textwidth]{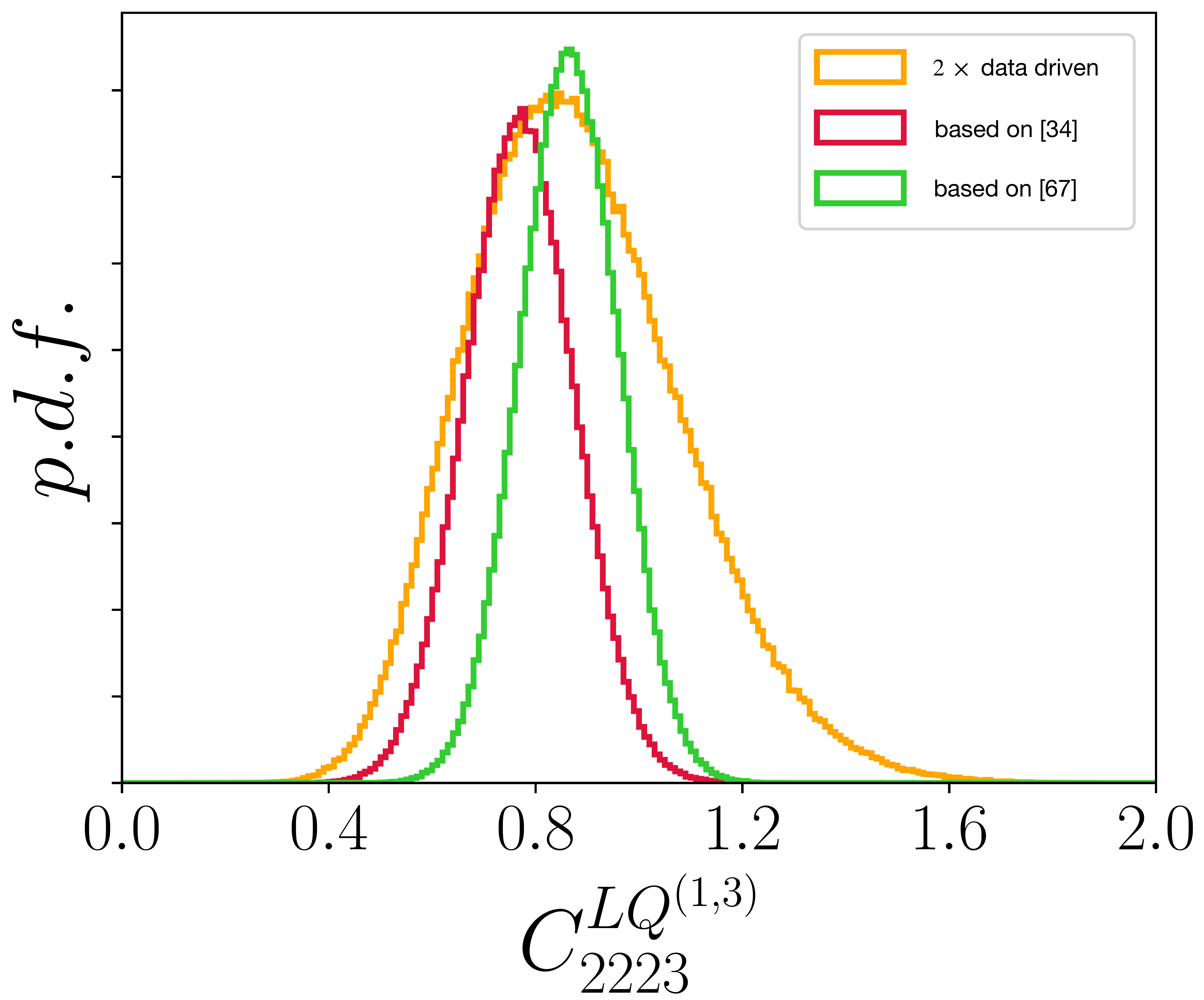}
\includegraphics[width=0.45\textwidth]{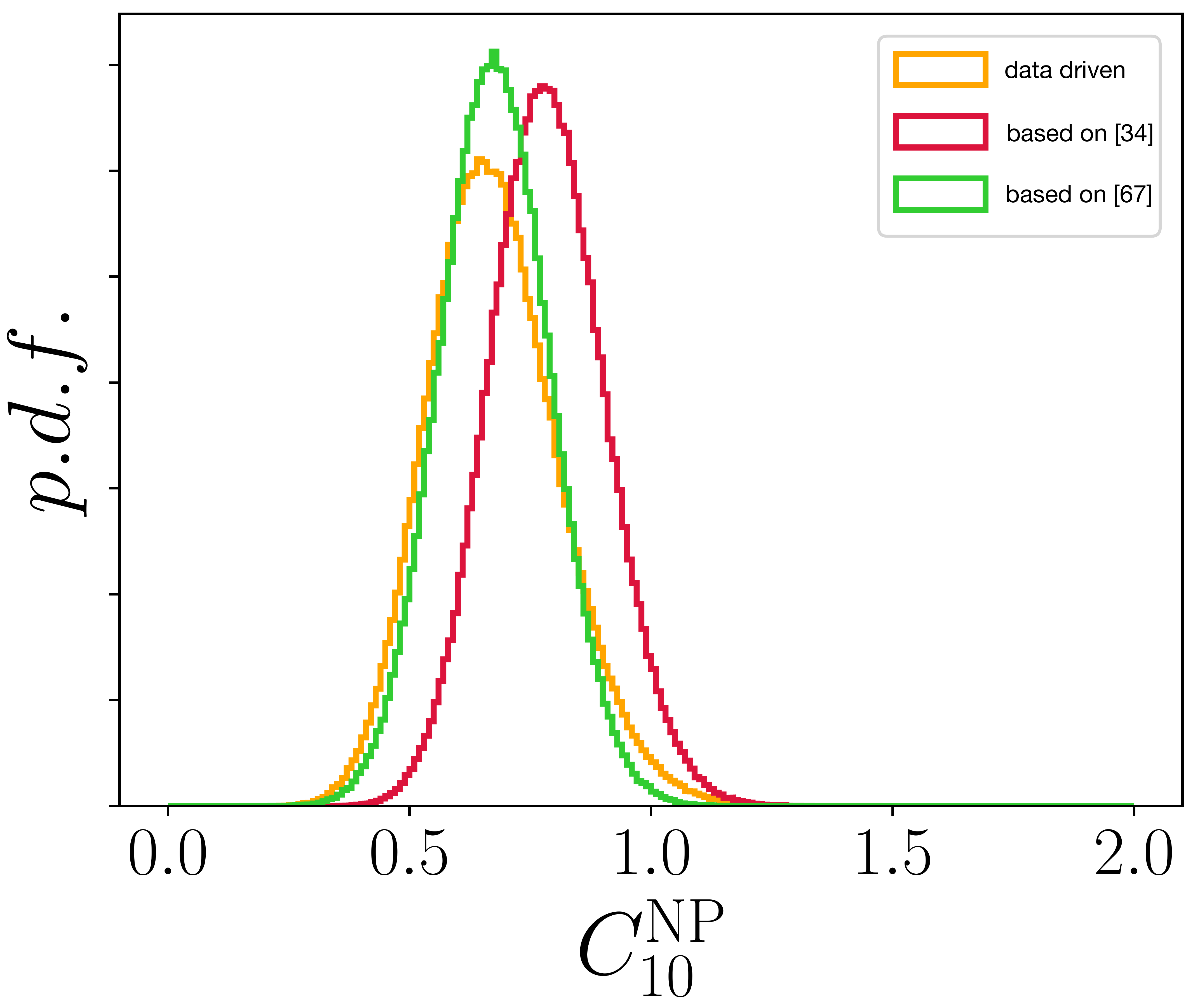}
\caption{Probability density function ({p.d.f.}) 
for $C_{9}^{\rm NP}$ (first panel), $C_{2223}^{LQ}$ (second panel) and $C_{10}^{\rm NP}$ (third panel). 
Green, red and orange {p.d.f.}'s correspond to the model-dependent approach from ref.~\cite{Khodjamirian:2010vf}, to the one from ref.~\cite{Ciuchini:2018anp}, and to the data-driven approach. }
\label{fig:1Dres}
\end{figure}

\begin{figure}[htb!]
\centering
\includegraphics[width=0.49\textwidth]{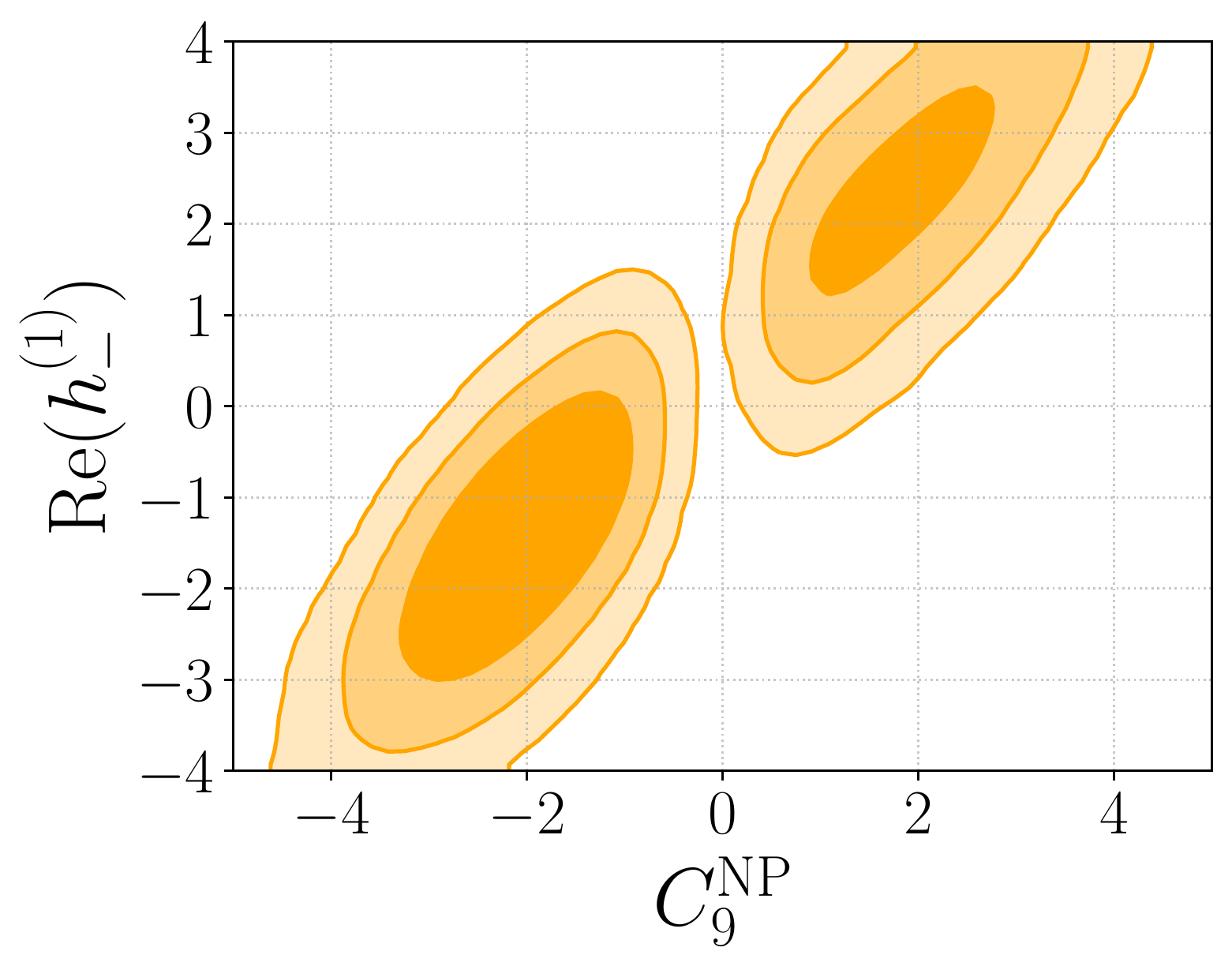}
\includegraphics[width=0.49\textwidth]{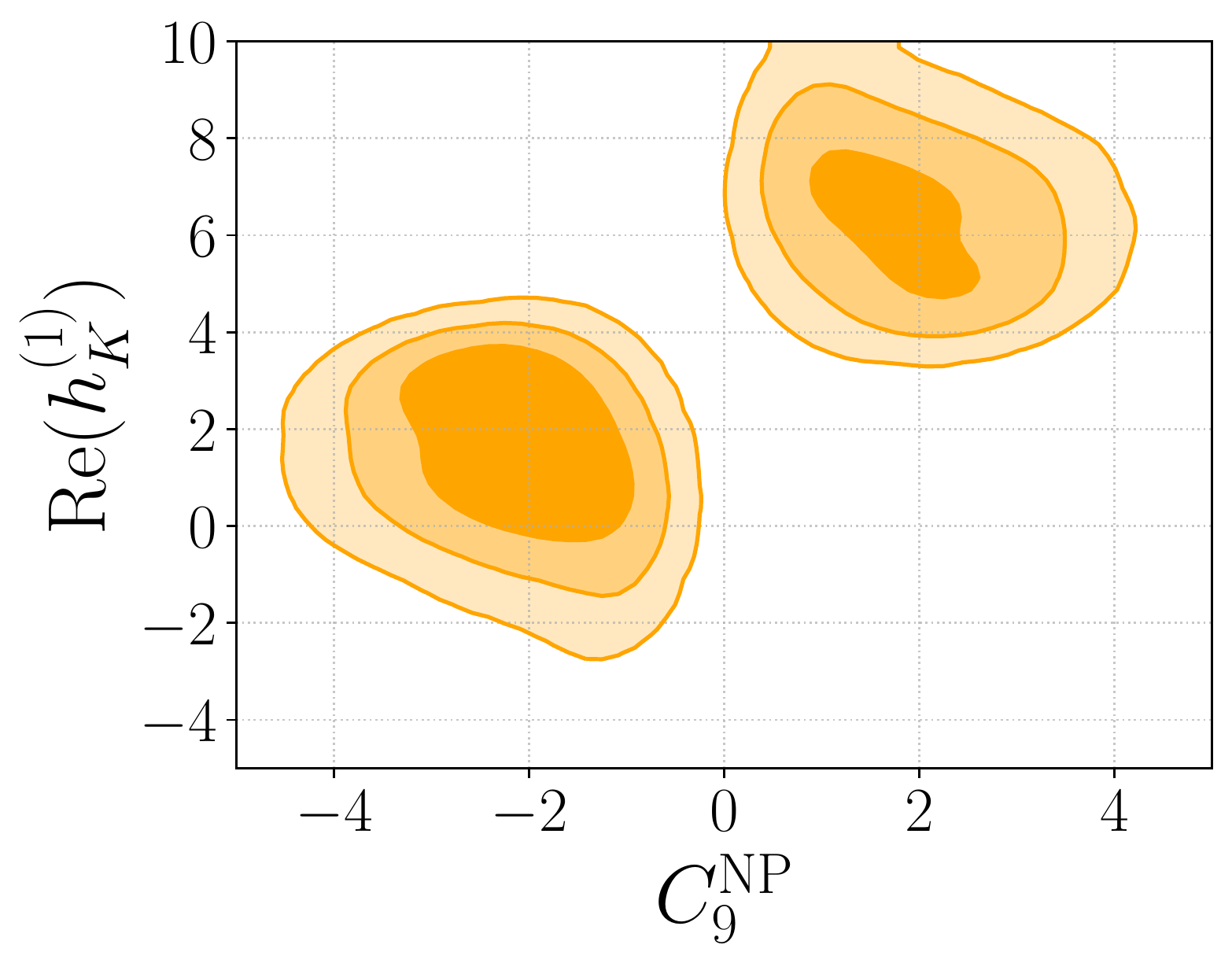}
\caption{Posteriors in the $(C_{9}^{\rm NP},\mathrm{Re}\,h_-^{(1)})$ plane for $B\to K^* \ell^+ \ell^-$ 
(left panel) and in the $(C_{9}^{\rm NP},\mathrm{Re}\,h_K^{(1)})$ plane for $B\to K \ell^+ \ell^-$ 
(right panel). The colour scheme is defined in the caption of Fig.~\ref{fig:1Dres}. Contours correspond to smallest regions of 68.3\%, 95.4\% and 99.7\% probability.}
\label{fig:C9hm}
\end{figure}

\begin{figure}[htb!]
\centering
\includegraphics[width=0.49\textwidth]{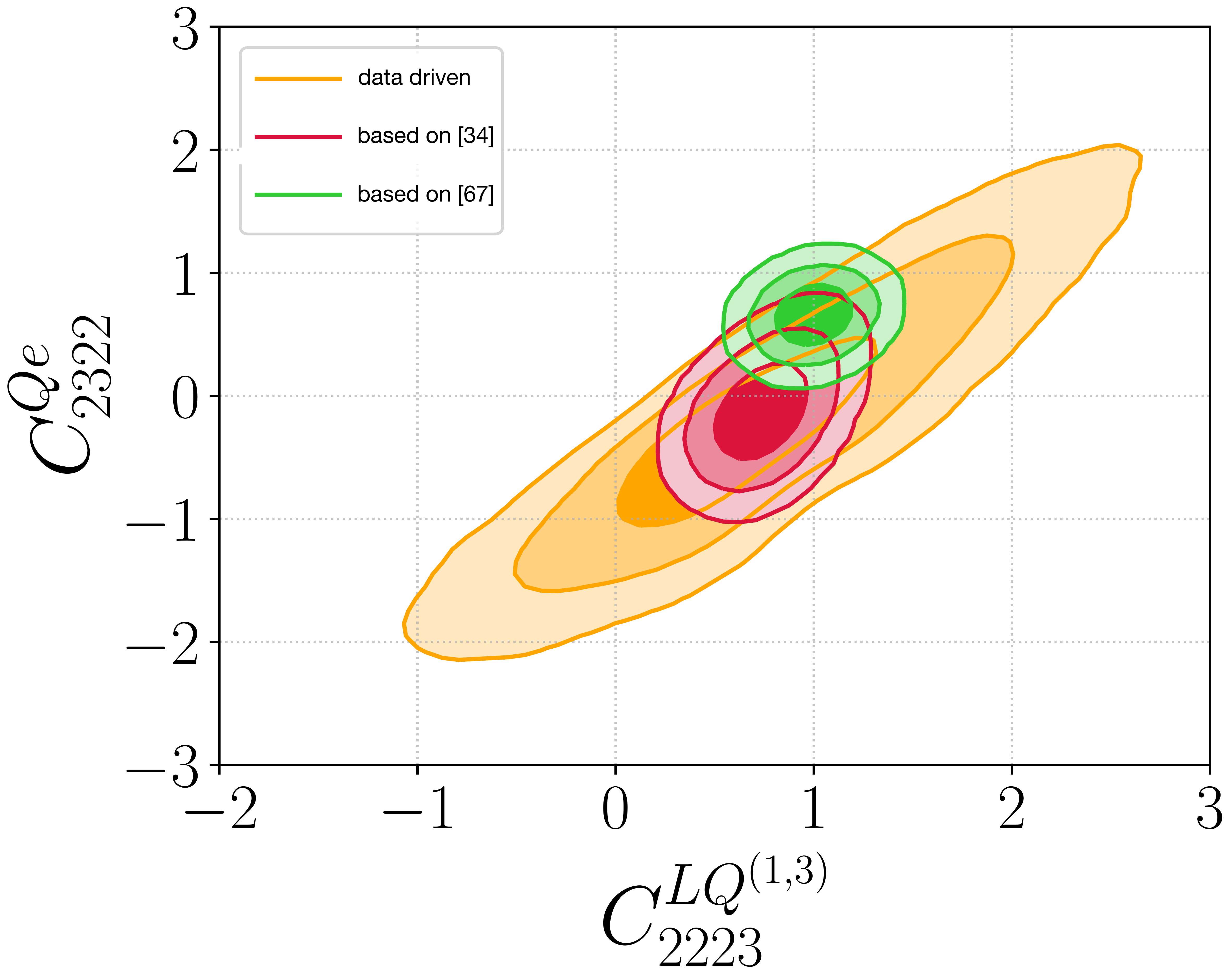}
\includegraphics[width=0.49\textwidth]{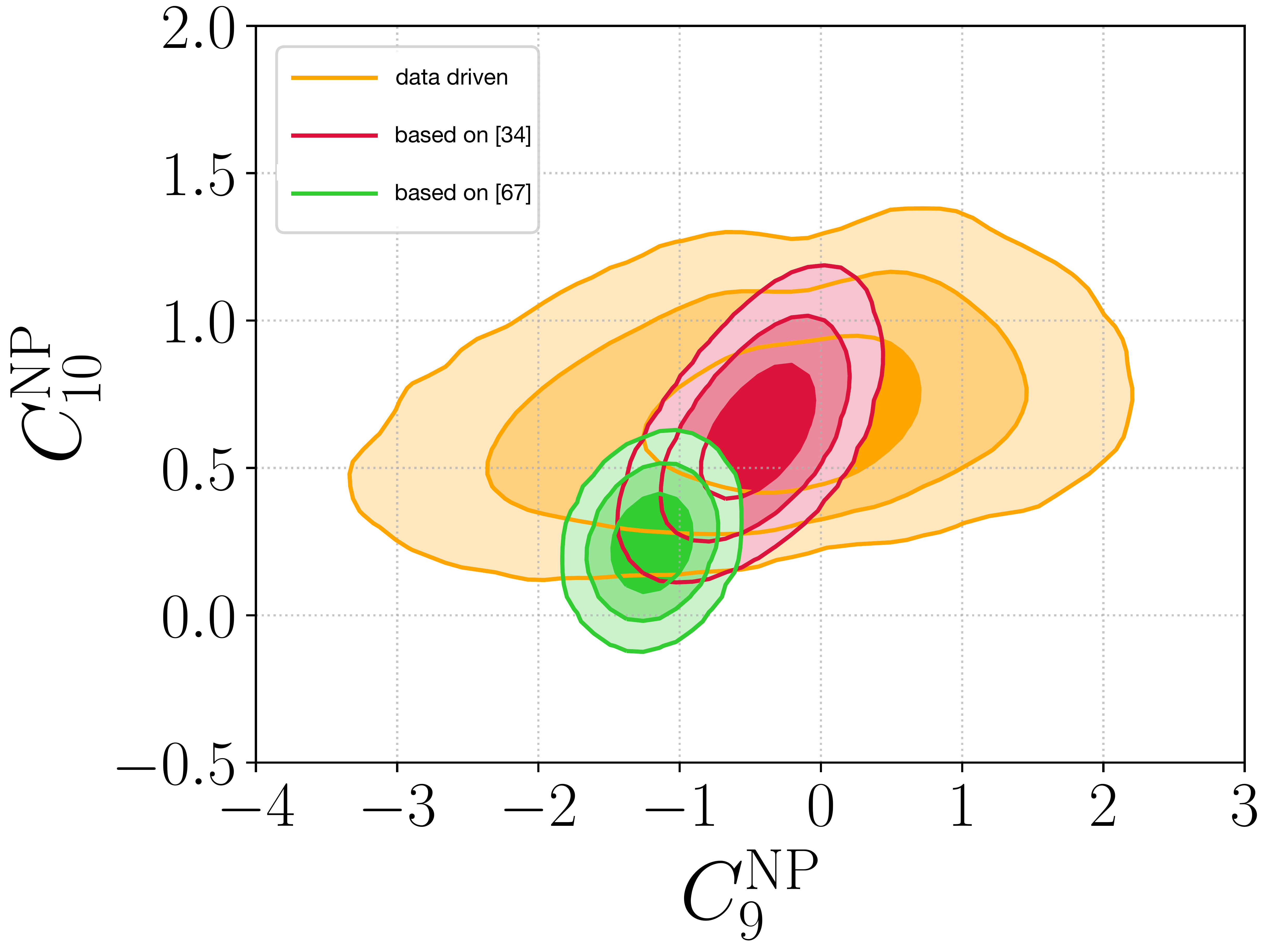}
\caption{Posteriors in the $(C_{2223}^{LQ},C_{2322}^{Qe})$  plane
(first panel) and in the $(C_{9}^{\rm NP},C_{10}^{\rm NP})$ plane (second panel). The colour scheme is defined in the caption of Fig.~\ref{fig:1Dres}. Contours correspond to smallest regions of 68.3\%, 95.4\% and 99.7\% probability. }
\label{fig:2Dres}
\end{figure}

We perform a Bayesian fit to the data in refs.~\cite{LHCb:2021trn,LHCb:2021lvy,Aaij:2017vbb,Abdesselam:2019wac,CMS:2014xfa,LHCb:2017rmj,ATLAS:2018cur,CMS:2019bbr,LHCb:2021vsc,Aaij:2020ruw,Aaij:2020nrf,Aaij:2015oid,LHCb:2013tgx,LHCb:2015wdu,LHCb:2021zwz,LHCb:2021xxq,Belle:2016fev,Belle:2019oag} in several NP scenarios characterized by different combinations of nonvanishing Wilson coefficients. To perform model comparison of different scenarios, we compute the \textit{Information Criterion} (IC) \cite{IC}: 
\begin{equation}\label{eq:IC}
   IC \equiv -2 \overline{\log \mathcal{L}} \, + \, 4 \sigma^{2}_{\log \mathcal{L}} \,,
\end{equation}
where the first and second terms represent mean and variance of the log likelihood posterior distribution, respectively. The first term measures the quality of the fit, while the second one counts effective degrees of freedom and thus penalizes more complicated models. Models with smaller IC should be preferred according to the canonical scale of evidence of Ref.~\cite{BayesFactors}, related in this context to (positive) IC differences. For convenience, we always report $\Delta IC \equiv IC_{\textrm{SM}} - IC_{\textrm{NP}}$.

As is evident from the discussion in \autoref{sec:Hadronic}, different assumptions on charming penguins yield different results on NP Wilson coefficients, since LUV ratios depend on the charm loop through the interference between NP and SM contributions. It goes without saying that a conservative inference on NP requires a conservative, i.e. data driven, estimate of charming penguins. Of course, since the SM reproduces much better experimental data in the data-driven approach (and also in the one based on ref.~\cite{Ciuchini:2018anp}) than in the approach from ref.~\cite{Khodjamirian:2010vf}, with the first scenario performing better than the second one, \textcolor{black}{the variation $\Delta IC$ due to NP (or equivalently the significance of NP) will be the smallest in the data-driven approach and the largest in the one based on ref.~\cite{Khodjamirian:2010vf}.} 

\begin{table*}
\centering
\renewcommand{\arraystretch}{1.5}
{\footnotesize
\centering
\begin{tabular}{|c|c|cccccc|}
\hline
& \textbf{Approach} & \boldmath$ R_K$ & \boldmath$ R_{K^*}$ & \boldmath$ R_{K^*}$
& \boldmath$ P_5'$ & \boldmath$ P_5'$ & \boldmath$B_s \to \mu\mu$ \\
&  & \boldmath$[1.1,6]$ & \boldmath$ [0.045,1.1]$ & \boldmath$ [1.1,6]$
& \boldmath$ [4,6]$ & \boldmath$ [6,8]$ & $\times 10^9$ \\
\hline
Exp. & -
& 0.848(42) & 0.680(93) & 0.71(10) & -0.439(117) & -0.583(095) & 2.86(33) \\
\hline
\multirow{3}{*}{A}
& data driven &
0.84(4)  & 0.86(4) & 0.81(13)  & -0.47(5)  & -0.53(7)  & 3.58(11)  \\ 
 & based on~\cite{Ciuchini:2018anp} &
0.76(4)  & 0.89(1) & 0.85(3)  & -0.44(5)  & -0.55(6)  & 3.58(11)  \\ 
 & based on~\cite{Khodjamirian:2010vf} & 
0.76(2)  & 0.89(1) & 0.83(1)  & -0.45(4)  & -0.59(4)  & 3.58(11)  \\ 
\hline
\multirow{3}{*}{B}
& data driven &
0.83(4)  & 0.85(2)  & 0.75(5)  & -0.48(5)  & -0.54(7)  & 2.64(21)  \\
 & based on~\cite{Ciuchini:2018anp} &
0.76(3)  & 0.86(1)  & 0.76(3)  & -0.46(5)  & -0.56(6)  & 2.74(11)  \\
 & based on~\cite{Khodjamirian:2010vf} &
0.72(3)  & 0.85(1)  & 0.74(3)  & -0.63(3)  & -0.74(2)  & 2.65(10)  \\
\hline
\multirow{3}{*}{C}
& data driven &
0.82(3)  & 0.86(1)  & 0.75(5)  & -0.49(5)  & -0.55(7)  & 2.56(19)  \\
 & based on~\cite{Ciuchini:2018anp} &
0.83(2)  & 0.85(1)  & 0.76(3)  & -0.48(5)  & -0.57(6)  & 2.40(16)  \\
 & based on~\cite{Khodjamirian:2010vf} &
0.84(3)  & 0.87(1)  & 0.74(3)  & -0.73(3)  & -0.80(2)  & 2.55(16)  \\
\hline
\multirow{3}{*}{D}
& data driven &
0.83(4)  & 0.85(2)  & 0.75(6)  & -0.49(5)  & -0.55(7)  & 2.58(23)  \\
 & based on~\cite{Ciuchini:2018anp} &
0.77(4)  & 0.85(1)  & 0.76(3)  & -0.47(5)  & -0.57(6)  & 2.67(21)  \\
 & based on~\cite{Khodjamirian:2010vf} &
0.71(3)  & 0.87(1)  & 0.77(3)  & -0.48(4)  & -0.62(4)  & 3.20(16)  \\
\hline
\multirow{3}{*}{E}
& data driven &  
0.84(4)  & 0.82(4)  & 0.68(8)  & -0.48(6)  & -0.55(7)  & 2.54(29)  \\
& based on~\cite{Ciuchini:2018anp} &
0.79(4)  & 0.81(3)  & 0.65(8)  & -0.47(5)  & -0.56(6)  & 2.64(24)  \\
 & based on~\cite{Khodjamirian:2010vf} &
0.80(4)  & 0.82(2)  & 0.67(4)  & -0.49(4)  & -0.64(4)  & 2.80(22)  \\
\hline
\end{tabular}
}
\caption{Experimental measurements with symmetrized errors (for $R_{K^*}$ and $P_5'$ we report the LHCb ones) and posteriors for key observables in the SMEFT scenarios considered here. Scenario A corresponds to NP contributions to $C_9$ only; scenario B to NP contributions to $C_{2223}^{LQ}$ only; scenario C to NP contributions to $C_{10}$ only; scenario D to NP contributions to $C_{2223}^{LQ}$ and $C_{2322}^{Qe}$; and scenario E to NP contributions to $C_{2223}^{LQ}, C_{2322}^{Qe}, C_{2223}^{Ld}$ and $C_{2223}^{ed}$. \label{tab:OBS_SMEFT}}
\end{table*}

Let us first consider three very simple NP scenarios: scenario A, in which deviations can only arise in $C_9$, scenario B, in which only $C^{LQ}_{2223}$ can be nonvanishing, corresponding to $C_{9}^\mathrm{NP} = - C_{10}^\mathrm{NP}$, and scenario C, in which only $C_{10}^\mathrm{NP}$ is allowed to float. Already in these simple NP scenarios there are dramatic differences in the fit depending on the assumption on charming penguins. Under the data-driven approach (and the one based on ref.~\cite{Ciuchini:2018anp}), scenarios B and C perform much better than scenario A, while the opposite is true in the approach from ref.~\cite{Khodjamirian:2010vf}. As reported in the \textcolor{black}{top} panel of Fig.~\ref{fig:1Dres}, in the data-driven approach charming penguins can even interfere destructively with $C_{9}^\mathrm{NP}$, allowing for a second solution for LUV observables with positive $C_{9}^\mathrm{NP}$, albeit with a smaller $\Delta IC$ with respect to the solution with negative $C_{9}^\mathrm{NP}$ (see Table \ref{tab:WC_SMEFT})\footnote{\textcolor{black}{Given the large dimensionality of the problem, we adopted in this study Metropolis-Hastings sampling. To handle multi-}\textcolor{black}{modal distributions we have generated a large number of chains ($\sim 500$) and checked the stability of the relative weight of the two modes under variations of the number of chains.}}.
{The correlation of $C_{9}^\mathrm{NP}$ with the hadronic parameters Re $h^{(1)}_{-}$ for final states involving $K^*$ and $K$ mesons is shown in Figure \ref{fig:C9hm}.} In the approach based on ref.~\cite{Khodjamirian:2010vf}, scenario A is ideal since it allows to strongly improve the agreement of both LUV and angular observables, while in scenario B the constraint from $B_s \to \mu^+ \mu^-$ limits the improvement in angular observables (see Table \ref{tab:OBS_SMEFT}), and scenario C cannot improve the agreement with angular observables at all. 
\textcolor{black}{Going back to Fig.~\ref{fig:P5p}, note that the global fit of scenario A would be perfectly in line with the hadronic contributions estimated in refs.~\cite{Khodjamirian:2010vf,Gubernari:2020eft}, related to what reported in green in the figure; while for scenario B and, most importantly, for C, the prediction for the observable $P_{5}^{\prime}$ would confront well with data only in the instance of sizable charming-penguin amplitudes.}
Conversely, under the data driven hypothesis, scenarios B and C allow to reproduce all observables, including LUV and $B_s \to \mu^+ \mu^-$, 
and therefore stand out as the preferred NP scenarios. The case where hadronic uncertainties are treated as in ref.~\cite{Ciuchini:2018anp} is in a somewhat intermediate position, with scenario C somewhat disfavoured with respect to scenario B due to the constraints on the charming penguin at low $q^2$. Obviously, as can be seen in Fig.~\ref{fig:1Dres}, the p.d.f. for $C_{10}^\mathrm{NP}$ in scenario C is almost independent of the hadronic uncertainties, while the overall quality of the fit strongly depends on the charming penguins, since in this scenario one needs hadronic contributions to reproduce the angular distributions and BRs.

\begin{figure*}[t!]
\includegraphics[width=0.99\textwidth]{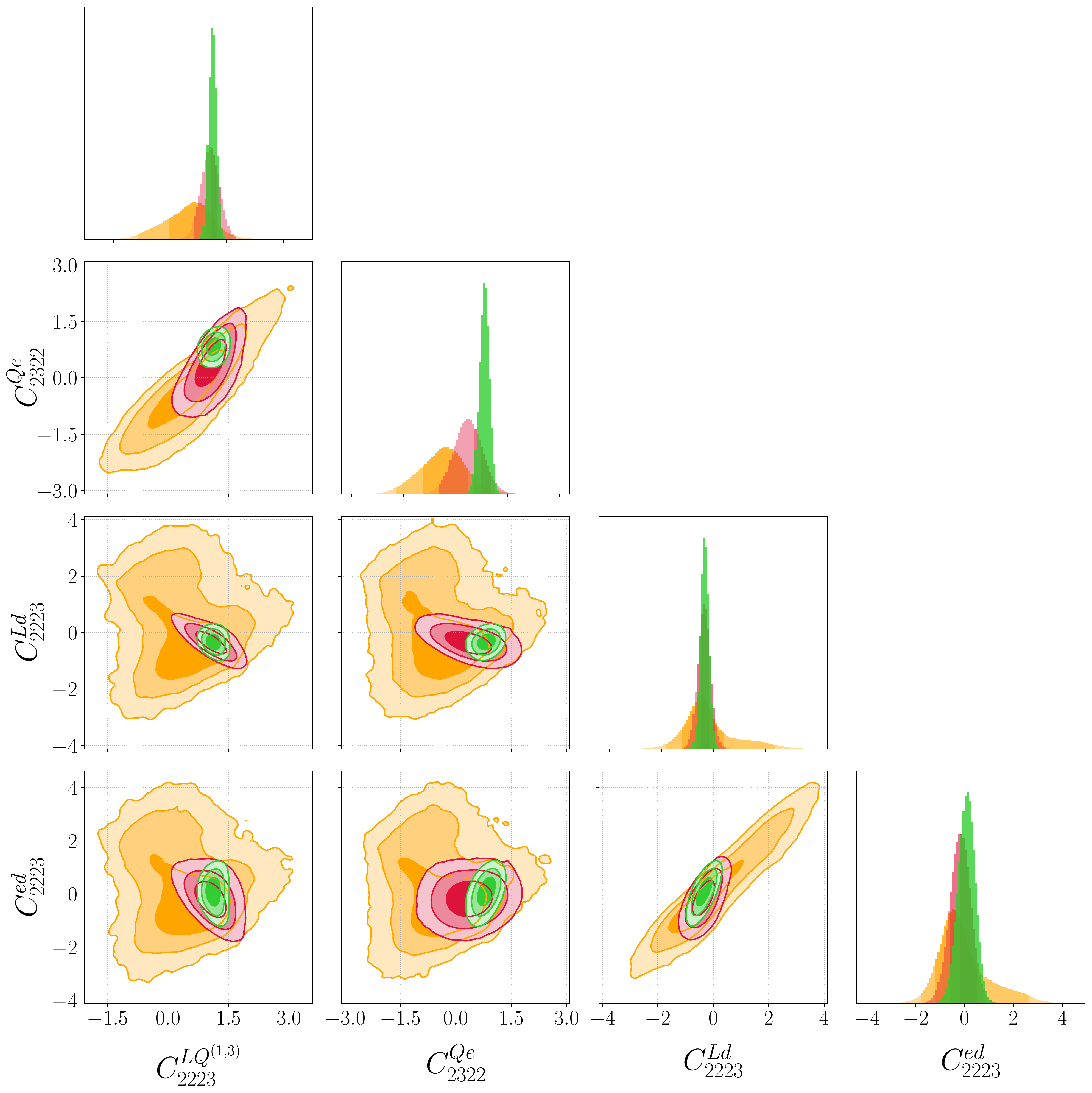}
\caption{Posteriors for $C_{2223}^{LQ}$, $C_{2322}^{Qe}$, $C^{Ld}_{2223}$ and $C^{ed}_{2223}$. Contours and colours as in Fig.~\ref{fig:2Dres}.}
\label{fig:4Dres_SMEFT}
\end{figure*}

\begin{figure*}[!ht!]
\includegraphics[width=0.99\textwidth]{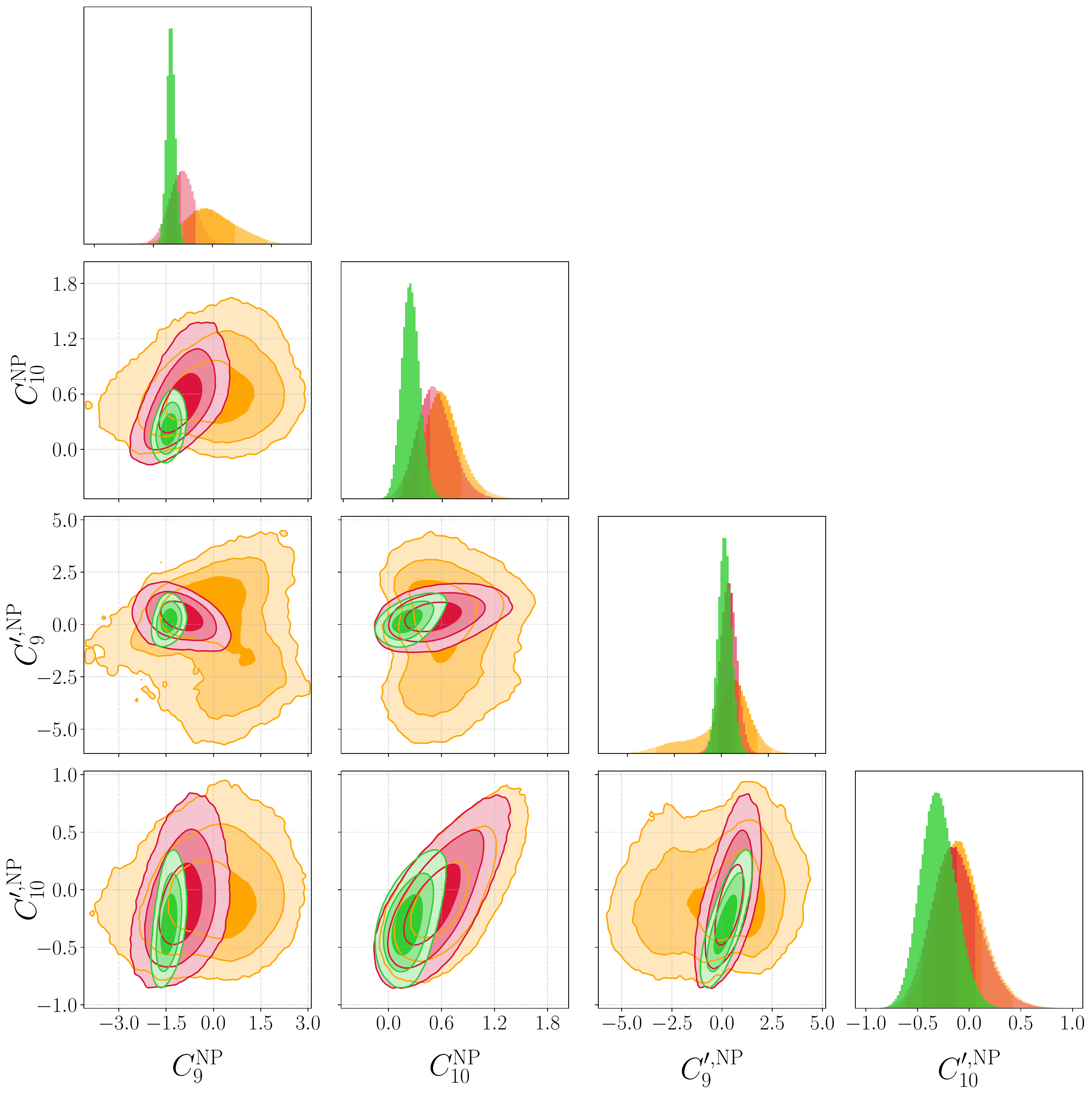}
\caption{Posteriors for$C_{9}^{\rm NP},C_{10}^{\rm NP}, C_{9}^{\prime, \rm NP}$ and $C_{10}^{\prime, \rm NP}$. Contours and colours as in Fig.~\ref{fig:2Dres}.}
\label{fig:4Dres_WEFT}
\end{figure*}

More general scenarios with two or more nonvanishing NP Wilson coefficients, such as scenario D, where $C^{LQ}_{2223}$ and $C^{Qe}_{2322}$ are allowed to float, or scenario E, where all the coefficients of the operators in eq.~(\ref{eq:SMEFT_op_tree}) are turned on, are slightly penalized by the number of degrees of freedom unless the approach from ref.~\cite{Khodjamirian:2010vf} is considered, as can be seen from Table \ref{tab:WC_SMEFT}. For the reader's convenience, in Table \ref{tab:WC_SMEFT} and in Figs.~\ref{fig:2Dres}-\ref{fig:4Dres_WEFT} we present results for scenarios D and E also in the weak effective Hamiltonian basis through eq.~(\ref{eq:SMEFT_matching}). 

It is interesting to look at the shape of the probability density contours for the NP parameters in scenario D reported in Fig.~\ref{fig:2Dres}. In the data-driven approach, $C_9^\mathrm{NP}$ is well compatible with 0, while a nonvanishing NP axial lepton coupling emerges from the experimental information coming from ``clean observables''. A slight preference for a nonvanishing $C_9^\mathrm{NP}$ is present in the model-dependent approach of ref.~\cite{Ciuchini:2018anp}, while a strong evidence for $C_9^\mathrm{NP}$ is obtained in the one based on ref.~\cite{Khodjamirian:2010vf}, together with a slight hint of a nonvanishing $C_{10}^\mathrm{NP}$. Therefore, Fig.~\ref{fig:2Dres} represents a clear example of how the \textcolor{black}{choice of the parameterization for the} charming penguins can strongly impact the inference of the underlying NP picture, allowing to go from a purely axial NP coupling to a purely vectorial one in the two extreme cases, with dramatic consequences for the model building related to $B$ anomalies. 

Concerning scenario E, it is worth noticing that in the case of the approach from ref.~\cite{Khodjamirian:2010vf}, right-handed operators allow to improve the agreement with $R_K$, given the current experimental hint for $R_{K} \neq R_{K^{*}}$ at the 1$\sigma$ level, see the discussion in \cite{Ciuchini:2019usw}. In the data-driven approach (and in the one based on ref.~\cite{Ciuchini:2018anp}) this can be achieved also through the interplay of hadronic corrections with LUV NP (see Table \ref{tab:OBS_SMEFT}). See Figs.~\ref{fig:4Dres_SMEFT} and \ref{fig:4Dres_WEFT} for a comparison of the posteriors for NP coefficients in scenario E.

\section{Conclusions}
\label{sec:Concl}

We have presented a global analysis of the experimental data on $b \to s \ell^+ \ell^-$ transitions from refs.~\cite{LHCb:2021trn,LHCb:2021lvy,Aaij:2017vbb,Abdesselam:2019wac,CMS:2014xfa,LHCb:2017rmj,ATLAS:2018cur,CMS:2019bbr,LHCb:2021vsc,Aaij:2020ruw,Aaij:2020nrf,Aaij:2015oid,LHCb:2013tgx,LHCb:2015wdu,LHCb:2021zwz,LHCb:2021xxq,Belle:2016fev,Belle:2019oag} under three different assumptions about the size and shape of the charming penguin contribution: a data-driven approach, a model-dependent based on ref.~\cite{Ciuchini:2018anp} and a model-dependent one based on ref.~\cite{Khodjamirian:2010vf}. \textcolor{black}{We have shown how current data point to helicity and $q^2$ dependence of the charm loop, as evinced in red in  Table~\ref{tab:hlambda} from the HPDI of some of the key hadronic parameters investigated.} We have discussed the interplay of NP and hadronic contributions and the dependence of the inferred NP from the assumptions on the charm loop. 

More conservative hypotheses point to two simple NP scenarios, either a nonvanishing $C^{LQ}_{2223}$, with a $\Delta IC$ with respect to the SM of 33 (53) in the data-driven approach (in the one of ref.~\cite{Ciuchini:2018anp}) , or a nonvanishing $C_{10}^\mathrm{NP}$, 
with a $\Delta IC$ with respect to the SM of 34 (48) in the data driven case (in the one based on ref.~\cite{Ciuchini:2018anp}). The approach based on ref.~\cite{Khodjamirian:2010vf}, instead, favours a more complex scenario with four nonvanishing NP coefficients with a $\Delta IC$ of 98, although a $\Delta IC$ of 88 can be achieved in the simple scenario of a nonvanishing $C_9^{NP}$. Clearly, more data on both LUV observables and differential decay rates is needed to improve our understanding of the charm loop and to single out the correct interpretation of LUV in terms of NP contributions. Hopefully, the LHC \cite{Cerri:2018ypt,Aaij:2244311} and Belle II \cite{Kou:2018nap} will provide us with the needed precision to identify the NP at the origin of the current evidence of LUV.

\textbf{Acknowledgements.}The work of M.F. is supported by the Deutsche Forschungsgemeinschaft (DFG, German Research Foundation) under grant  396021762 - TRR 257, ``Particle Physics Phenomenology after the Higgs Discovery''. The work of M.V. is supported by the Simons Foundation under the Simons Bridge for Postdoctoral Fellowships at SCGP and YITP, award number 815892. The work of A.P. is funded by Volkswagen Foundation within the initiative ``Corona Crisis and Beyond -- Perspectives for Science, Scholarship and Society'', grant number 99091. This work was supported by the Italian Ministry of Research (MIUR) under grant PRIN 20172LNEEZ. This research was supported in part through the Maxwell computational resources operated at DESY, Hamburg, Germany.

\bibliographystyle{JHEP-CONF}
\bibliography{hepbiblio}

\end{document}